\documentclass[epjc3,smallextended,final,twocolumn]{svjour3}

\usepackage{cite} 
\usepackage[english]{babel} 
\usepackage[utf8]{inputenc}
\usepackage{latexsym} 
\usepackage{amsmath} 
\usepackage{amsfonts} 
\usepackage{amssymb} 
\usepackage{textcomp} 
\usepackage{graphicx}  
\usepackage{enumerate} 
\usepackage{array} 
\usepackage{mathrsfs}
\usepackage{bm,bbm}
\usepackage{bbold}  
\usepackage{multirow}
\usepackage{fancyref}
\usepackage[breaklinks]{hyperref}
\usepackage{color}
\usepackage{mathtools}

\newcommand{\be}{\begin{equation}}
\newcommand{\ee}{\end{equation}}
\newcommand{\ba}{\begin{eqnarray}}
\newcommand{\ea}{\end{eqnarray}}
\def\bs{\begin{subequations}}
\def\es{\end{subequations}}
\renewcommand{\leq}{\leqslant}
\renewcommand{\geq}{\geqslant}
\def\a{\alpha}
\def\b{\beta}

\def\g{\gamma}

\def\om{\omega}

\def\cL{\mathcal{L}}

\def\p{\partial}

\newcommand{\Eq}[1]{(\ref{#1})}
\def\com{\color{magenta}}
\def\cob{\color{blue}}
\def\Pl{{\rm Pl}}
\newcommand{\oarX}[1]{\href{http://arxiv.org/abs/#1}{{\ttfamily\com arXiv:#1}}}
\newcommand{\arX}[1]{\href{http://arxiv.org/abs/#1}{{\ttfamily\com arXiv:#1}}}

\newcommand{\doin}[6]{\href{http://dx.doi.org/#1}{\cob  #2 #3 {\bf #4}, #5 (#6)}}
\newcommand{\doinn}[5]{\href{http://dx.doi.org/#1}{\cob  #2 {\bf #3}, #4 (#5)}}
\newcommand{\doij}[5]{\href{http://dx.doi.org/#1}{\cob  #2 {\bf #3}, #4 (#5)}}

\newcommand{\book}[5]{{\it #1} (#2, #3, #5)}
\newcommand{\books}[4]{{\it #1} (#2, #3, #4)}
\newcommand{\proc}[6]{In \emph{#1}, ed.\ by #2 (#3, #4, #6)}
\newcommand{\tia}[1]{#1.}
\newcommand{\tiaq}[1]{#1}

\newcommand{\bea}{\begin{eqnarray}}
\newcommand{\eea}{\end{eqnarray}}

\def\rmd{\mathrm{d}}

\journalname{Eur. Phys. J. C}

\date{March 22, 2017}

\begin{document}\sloppy

\title{Black holes in multi-fractional and Lorentz-violating models}

\author{Gianluca Calcagni\thanksref{addr1,e1} \and David Rodr\'iguez Fern\'andez\thanksref{addr2,e2} \and Michele Ronco\thanksref{addr3,addr4,e3}}

\thankstext{e1}{e-mail: calcagni@iem.cfmac.csic.es}
\thankstext{e2}{e-mail: rodriguezferdavid@uniovi.es}
\thankstext{e3}{e-mail: michele.ronco@roma1.infn.it}

\institute{Instituto de Estructura de la Materia, CSIC, Serrano 121, 28006 Madrid, Spain\label{addr1}
\and \noindent Department of Physics, Universidad de Oviedo Avda.\ Calvo Sotelo 18, ES-33007 Oviedo, Spain\label{addr2} \and \noindent Dipartimento di Fisica, Universit\`a di Roma ``La Sapienza'', P.le A.\ Moro 2, 00185 Rome, Italy\label{addr3} \and \noindent INFN, Sez.~Roma1, P.le A.\ Moro 2, 00185 Rome, Italy\label{addr4}}

\maketitle

\begin{abstract}
We study static and radially symmetric black holes in the multi-fractional theories of gravity with $q$-derivatives and with weighted derivatives, frameworks where the spacetime dimension varies with the probed scale and geometry is characterized by at least one fundamental length $\ell_*$. In the $q$-derivatives scenario, one finds a tiny shift of the event horizon. Schwarzschild black holes can present an additional ring singularity, not present in general relativity, whose radius is proportional to $\ell_*$. In the multi-fractional theory with weighted derivatives, there is no such deformation, but non-trivial geometric features generate a cosmological-constant term, leading to a de Sitter--Schwarzschild black hole. For both scenarios, we compute the Hawking temperature and comment on the resulting black-hole thermodynamics. In the case with $q$-derivatives, black holes can be hotter than usual and possess an additional ring singularity, while in the case with weighted derivatives they have a de Sitter hair of purely geometric origin, which may lead to a solution of the cosmological constant problem similar to that in unimodular gravity. Finally, we compare our findings with other Lorentz-violating models.
\end{abstract}


\section{Introduction}

Along the winding road to quantum gravity, it is easy to stop by and get absorbed by any of the local views offered by the scenery we find when classical general relativity (GR) is abandoned and the territory of pre-geometry, modified gravity, discrete spacetimes, and all the rest, is entered. The question of how gravity is affected when it becomes quantum or is changed by phenomenological reasons receives different answers according to the scale of observation; cosmology, astrophysics, and even atomic physics can give complementary information on how matter and geometry behave when the principles of general relativity and quantum mechanics are unified or modified \cite{book,bojo1,Matt,gacLivRev,bojo2,qgAtom,rev}.

Among the most recent theories beyond Einstein gravity, \emph{multi-fractional spacetimes} \cite{rev,frc2,frc11,trtls,first} have received some obstinate attention due to their potential in giving a physical meaning to several concepts scattered in quantum gravity. In particular, not only they allowed one to control the change of spacetime dimensionality typical of all quantum gravities analytically, but they also recognized this feature as a treasure trove for phenomenology, since it leaves an imprint in observations at virtually all scales. The main idea is simple. Consider the usual $D$-dimensional action $S=\int d^Dx\,\sqrt{-g}\,\cL[\phi^i,\p]$ of some generic fields $\phi^i$, where $g$ is the determinant of the metric and $\p$ indicates that the Lagrangian density contains ordinary integer-order derivatives. In order to describe a matter and gravitational field theory on a spacetime with geometric properties changing with the scale, one alters the integro-differential structure such that both the measure $d^Dx\to d^Dq(x)$ and the derivatives $\p_\mu\to\mathbbm{D}_\mu$ acquire a scale dependence, i.e., they depend on a hierarchy of scales $\ell_1\equiv\ell_*,\ell_2,\dots$. Without any loss of generality at the phenomenological level \cite{rev,first}, it is sufficient to consider only one length scale $\ell_*$ (separating the infrared from the ultraviolet). The explicit functional form of the \emph{multi-scale} measure depends on the symmetries imposed but it is universal once this choice has been made. For instance, theories of multi-scale geometry where the measure $d^Dq(x)=\prod_\mu\rmd q^\mu(x^\mu)$ is factorizable in the coordinates are called \emph{multi-fractional} theories and the $D$ profiles $q^\mu(x^\mu)$ are determined uniquely (up to coefficients, as we will discuss below) only by assuming that the spacetime Hausdorff dimension changes ``slowly'' in the infrared \cite{rev,first}. Below we will give an explicit expression. Quite surprisingly, this result, known as second flow-equation theorem, yields exactly the same measure one would obtain by demanding the integration measure to represent a deterministic multi-fractal \cite{frc2}. There is more arbitrariness in the choice of symmetries of the Lagrangian, which leads to different multi-scale derivatives $\mathbbm{D}_\mu$ defining physically inequivalent theories. Of the three extant multi-fractional theories (with, respectively, weighted, $q$- and fractional derivatives) two of them (with $q$- and fractional derivatives) are very similar to each other and especially interesting for the ultraviolet behavior of their propagator. Although a power-counting argument fails to guarantee renormalizability, certain fractal properties of the geometry can modify the poles of traditional particle propagators into some fashion yet to be completely understood \cite{rev}. The same fractal properties can affect also the big-bang singularity, either removing it or altering its structure, as it might be the case for the theories with weighted and $q$-derivatives \cite{rev}. Since the fate of singularities is an important element to take into consideration when assessing alternative theories of gravity or particle physics, the next obvious step is to check what happens for black holes in multi-fractional spacetimes. 

This issue has not been tackled before and is the goal of this paper. We study static\footnote{We take the Schwarzschild solution because it is the simplest one that describes a black hole.} and spherically symmetric black-hole solutions in two different multi-fractional theories, with $q$-derivatives and with weighted derivatives. In both cases, the background-independent gravitational action has been known for some time but only cosmological solutions have been considered so far \cite{frc11}. We find interesting departures from the Schwarzschild solution of GR. The size and topology of event horizons and singularities are indeed deformed by multi-fractional effects. Conceptually, the interest of this resides in the fact that it represents a top-down example where small-scale modifications of standard GR affect (even if in a tiny way) the physics on large scales, namely the structure of the Schwarzschild horizon. Of the two solutions found in the case with weighted derivatives, in one there is no deformation on the horizon radius, while in the other there is. The Hawking temperature (hence, in principle, also black-hole thermodynamics) is modified in both multi-fractional theories. 

In all the cases, we restrict ourselves to small deformations due to anomalous effects, consistently with observational bounds on the scales of the geometry \cite{rev}. Because of this, and as confirmed through computations, all the predictions we make (for instance, deviations in the evaporation time of black holes) correspond to tiny deviations with respect to the standard framework.

Our paper is organized as follows. In Sect.\ \ref{II}, we analyze static and spherically symmetric black-hole solutions in multi-fractional gravity with $q$-derivatives. Having reviewed briefly the latter, we give a self-consistent discussion on the presentation issue in Sect.~\ref{IIA}. In Sects.\ \ref{IIBa}--\ref{IIBc}, we study the properties of $q$-multi-fractional black holes, focusing on the event horizon, the curvature singularity and the Hawking temperature. All these pivotal features of general-relativistic black holes are deformed by multi-scaling effects. In Sect.\ \ref{IIC}, we notice that, for certain choices of measure and presentation, quantum modifications of the ergosphere combined with gravitational-wave data can efficiently constrain the multi-scale length $\ell_*$. Black holes in multi-fractional gravity with weighted derivatives are analyzed in Sect.\ \ref{III} and we find that they are standard Schwarzschild black holes with a cosmological constant term. The specific form of the solution depends on a parameter $\Omega$, related to the kinetic term for the measure profile. In Sect.\ \ref{IIIA}, we focus on the solution for $\Omega=0$. Then, we analyze how the evaporation time is slightly modified by multi-scale effects in Sect.\ \ref{IIIB}. The simplest case in which $\Omega \neq 0$ is studied in Sect.~\ref{IIIC}, showing in \ref{IIID} that the consequences on the evaporation time are also of the same order. In Sect.\ \ref{IV}, we discuss similarities and differences between our results and black holes in Lorentz-violating theories of gravity, such as Ho\v{r}ava–-Lifshitz gravity, non-commutative and non-local models. Finally, in Sect.\ \ref{V} we summarize our results and outline possible future plans.

Throughout the paper, we work with $k_{\rm B} = \hbar \, = \, c = 1$, if not specified differently. 


\subsection{A note on terminology and fractional calculus}

Multi-fractional theories propose an extension of certain aspects of fractional calculus to a multi-dimensional ($D>1$ topological dimensions) multi-scale (scale-dependent scaling laws) setting. In this paper, we will not need the elegant tools of fractional calculus \cite{SKM,GoMa,KST,Die10,frc1} but some clarification of terminology may be useful. In all multi-fractional theories, the ultraviolet part of the factorizable integration measure $\prod_\mu\rmd q^\mu(x^\mu)$ is, for each direction, the measure of fractional integrals, as discussed at length in \cite{frc1}. The only (but important) detail that changes with respect to the fractional integrals defined in the literature \cite{SKM,GoMa,KST,Die10} is the support of the measure, in this case the whole space \cite{rev,frc2,frc1} instead of the half line. The scaling property is the same as in fractional integrals and one can consider the multi-fractional measure as a proper multi-scale, multi-dimensional extension\footnote{A multi-dimensional extension of fractional operators in classical mechanics was first considered in \cite{ElT2}.} of the latter. Hence the name of this class of theories.

On the other hand, only the theory dubbed $T_\g$ \cite{rev} features (the multi-scale extension \cite{rev,frc2} of) fractional derivatives and we will not employ this naming for anything else. The models discussed in this paper have other types of operators, called $q$-derivatives and weighted derivatives, neither of which is fractional. In particular, $q$-derivatives may be regarded as an approximation of fractional derivatives \cite[question {\it 13}]{rev} but they are much simpler than them. In \cite{frc2}, the name ``$q$-derivative'' was inspired by the coordinate labels in classical mechanics and was consistently used in subsequent papers to indicate a specific operator we will discuss later. Unfortunately, this is the same name of another, quite different operator introduced by Jackson in 1909 \cite{Jac09} and utilized in Tsallis thermodynamics \cite{Tsa09}. The difference in context will avoid confusion between our $q$-derivatives and Jackson derivatives.


\section{Schwarzschild solution in the multi-fractional theory with \texorpdfstring{$q$}{}-derivatives} \label{II}

In the multi-fractional theory with $q$-derivatives, gravity is rather straightforward to work out \cite{frc11}. In fact, one only has to make the substitution $x^\mu \rightarrow q^\mu(x^\mu)$ everywhere in the action. In other words, ordinary derivatives $\partial_\mu$ are replaced by $[1/v_\mu(x^\mu)]\partial/\partial x^\mu=v_\mu^{-1}(x^\mu)\partial_\mu$ (no index summation), where $v_\mu=\p_\mu q^\mu$. As a result,  the Riemann tensor in this theory is
\begin{equation}
\label{riemten}
\prescript{q}{}{R}^\rho_{\mu \sigma \nu} = \frac{1}{v_\sigma}\partial_\sigma \prescript{q}{}{\Gamma}^\rho_{\mu\nu}-\frac{1}{v_\nu}\partial_\nu \prescript{q}{}{ \Gamma}^\rho_{\mu\sigma} + \prescript{q}{}{\Gamma}^\tau_{\mu\nu} \prescript{q}{}{\Gamma}^\rho_{\sigma\tau}-\prescript{q}{}{\Gamma}^\tau_{\mu\sigma}\prescript{q}{}{\Gamma}^{\rho}_{\nu\tau},
\end{equation} 
where the Christoffel symbol is
\begin{equation}
\label{connect}
\prescript{q}{}{\Gamma}^\rho_{\mu\nu} = \frac{1}{2}g^{\rho\sigma}\left(\frac{1}{v_\mu}\partial_\mu g_{\nu\sigma} + \frac{1}{v_\nu}\partial_\nu g_{\mu\sigma}-\frac{1}{v_\sigma}\partial_\sigma g_{\mu\nu}\right).
\end{equation}
Finally, the $q$-version of the Einstein--Hilbert action reads
\begin{equation}
\label{einact}
\prescript{q}{}{S} = \frac{1}{2 \kappa^2} \int d^D x v(x) \sqrt{-g}(\prescript{q}{}{R} - 2 \Lambda) + S_{\rm m}\,,
\end{equation}
where $v(x)=\prod_\mu v_\mu(x^\mu)$ and $S_{\rm m}$ denotes the matter action.

As the reader has certainly noticed, there is no difference between GR and multi-fractional gravity with $q$-derivatives when we write the latter in terms of $q^\mu$ coordinates. In fact, the geometric coordinates $q^\mu$ provide a useful way of re-writing the theory in such a way that all non-trivial aspects are hidden. However, the operation we described as ``$x^\mu\to q^\mu(x^\mu)$'' is only a convenient way of writing this theory from GR and it should not be confused with a standard coordinate change trivially mapping the physical dynamics onto itself. The presence of a background scale dependence (a structure independent of the metric and encoded fully in the profiles $q^\mu(x^\mu)$, which will be given {a priori}) introduces a preferred frame (called fractional frame, labeled by the fractional coordinates $x^\mu$) where physical observables must be calculated. In the fractional frame, where the integration measure gets non-trivial contributions $d^D x\, v(x)=d^Dx(1+\dots)$ and derivatives are modified into operators $v_\mu^{-1}(x^\mu) \partial_\mu$, one sees departures from GR. This point will be further explained and discussed in Sect.\ \ref{IIA}.

In the light of Eqs.\ \eqref{riemten}--\eqref{einact}, it is not difficult to realize that the solutions to Einstein equations are the same of GR  when they are expressed in $q^\mu$ coordinates, but non-linear modifications appear when we rewrite the solution as a function of $x^\mu$ by using the profiles $q^\mu(x^\mu)$. In the first part of this work, we shall show that these multi-fractional modifications affect not only the event horizon and the curvature singularity but also thermodynamic properties of black holes such as the Hawking temperature. 

After having reviewed the issue of presentation in Sect.\ \ref{IIA}, we shall show in Sect.\ \ref{IIBa} that the position of the horizon of the Schwarzschild black hole is generally shifted in multi-fractional gravity with $q$-derivatives. Depending on the choice of the presentation and also on the way we interpret the existence of a presentation ambiguity, the curvature singularity can remain unaffected or an additional singularity (or even many additional singularities, if we take into account {logarithmic oscillations}; see below) can appear, or there can be a sort of quantum uncertainty in the singularity position (Sect.\ \ref{IIBb}). As we shall explain exhaustively in the following, the interpretation of the results depends on how we interpret the presentation ambiguity in multi-fractional theories. A  general feature is that those extra singularities we find are non-local because they have a ring topology. Evaporation and the quantum ergosphere are considered in Sects.\ \ref{IIBc} and \ref{IIC}, respectively.


\subsection{The choice of presentation}\label{IIA}

Before analyzing the Schwarzschild solution in the multi-fractional formulation of gravity with $q$-derivatives, we review the so-called problem of \textit{presentation}.  We refer to \cite{rev,trtls,minlen} for further details. In \cite{trtls}, the presentation is described as an ambiguity of the model we have to fix. In \cite{minlen}, an interesting reinterpretation of the presentation issue is proposed: it would be not an ambiguity to fix but, rather, a manifestation of a microscopic stochastic structure in multi-fractional spacetimes. Here we are going to allow for both possibilities \cite{rev}.

Geometric coordinates $q^\mu(x^\mu)$ correspond to rulers which adapt to the change of spacetime dimension taking place at different scales. However, our measuring devices have fixed extensions and are not as flexible. For this reason, physical observables have to be computed in the fractional frame, i.e., the one spanned by the scale-independent coordinates $x^\mu$. This poses the problem of fixing a fractional frame $x^\mu$, where we can make physical predictions. To say it in other words, while dynamics formulated in terms of geometric coordinates is invariant under $q$-Poincaré  transformations $q^\mu({x'}^\mu)=\Lambda_\nu^{\ \mu}q^\nu(x^\nu)+a^\mu$, physical observables are highly $x$-frame dependent since, being them calculated in the fractional picture, they do not enjoy $q$-Poincaré (nor ordinary Poincaré) symmetries.\footnote{See Ref.\ \cite{CR} for a recent discussion of these symmetries.} As a consequence, multi-fractional predictions for a given observable hold only in the selected fractional frame. Thus, the choice of the presentation corresponds to fix a physical frame. We shall also comment on another way of looking at the problem of the presentation, which has been proposed recently \cite{minlen}. According to this new perspective, which we call ``stochastic view'' in contrast with the ``deterministic view'' where one must make a frame choice, the presentation ambiguity has the physical interpretation of an intrinsic limitation on the determination of distances due to stochastic properties of some multi-fractional theories.\footnote{Such a way of interpreting the presentation issue might hold rigorously only for the multi-fractional theory with fractional derivatives (but this point is still under study). On the other hand, the multi-fractional theory with $q$-derivatives is known to be an approximation of the theory with fractional derivatives in the infrared \cite{rev}, which guarantees that it also admits the stochastic view when not considered \emph{per se} \cite{minlen}.} In this paper, we shall follow both possibilities and underline how the interpretation of the results change according to the view we adopt.   

With the aim of defining fractional spherical coordinates (and, in particular, the fractional radius), we are particularly interested in analyzing the effect of different presentation choices on the definition of the distance in multi-fractional theories. We will define a radial coordinate sensitive to the presentation. In the $q$-theory in the deterministic view, there are only two choices available, each representing a separate model. On the other hand, according to the stochastic interpretation of the presentation ambiguity \cite{minlen}, there is no such proliferation of models because the two different available choices give the extremes of quantum uncertainty fluctuations of the radius.

We begin by recalling that, in the theory with $q$-derivatives, there is a specific relation between the distance $\varDelta x=x_{\rm B}-x_{\rm A}>0$ measured in the fractional frame and the (unphysical) geometric distance $\varDelta q>0$ (positive in order to have a well-defined norm). To be more concrete, let us consider the binomial measure
\be\label{mea}
q^\mu(x^\mu) = x^\mu +\frac{\ell_*}{\alpha}{\rm sgn}(x^\mu)\left |\frac{x^\mu}{\ell_*}\right|^\alpha.
\ee
This is the simplest measure entailing a varying dimension with the probed scale and is a very effective first-order coarse-graining approximation of the most general multi-fractional measure \cite{rev,first}. Let $D=1$ for the sake of the argument. Changing the presentation corresponds to making a translation: $q(x) \rightarrow \overline{q}(x) = q(x-\overline{x})$. The second flow-equation theorem \cite{first} fixes the possible values to $\overline{x}=x_{\rm A},x_{\rm B}$ \cite{rev} (corresponding to, respectively, the so-called initial-point and final-point presentation), notably excluding $\overline{x}=0$. Then the geometric distance $\varDelta q (x) =|\overline{q}(x_{\rm B})-\overline{q}(x_{\rm A})|$ is
\begin{equation}\label{quanpr}
\varDelta q(x) = \left| \varDelta x \pm \frac{\ell_*}{\alpha} \left|\frac{\varDelta x}{\ell_*}\right|^\alpha\right|,
\end{equation}
where the sign depends on the presentation choice ($+$, initial-point presentation; $-$, final-point presentation). As mentioned, there is also a different way of interpreting the presentation issue, motivated in \cite{minlen}. The multi-fractional correction to spacetime intervals may be regarded as a sort of uncertainty on the measurement, so that the two presentations in \Eq{quanpr} correspond to a positive or a negative fluctuation of the distance. From this point of view, we do not have to choose any presentation at all, since such an ambiguity has the physical meaning of a stochastic uncertainty. Interestingly, limitations on the measurability of distances can easily be obtained also by combining basic GR and quantum-mechanics arguments, thereby confirming that multi-fractional models encode semi-classical quantum-gravity effects \cite{minlen}.\footnote{For this reason, and with a slight abuse of terminology, we will interchangeably call these fluctuations of the geometry ``stochastic'' or ``quantum.''} Strikingly, this also provides an explanation for the universality of dimensional flow in quantum-gravity approaches: its origin is the combination of very basic GR and quantum-mechanics features. 

In this section, we are interested in studying the Schwarzschild solution in the multi-fractional theory with $q$-derivatives. To this aim, we first have to transform the multi-fractional measure to spherical coordinates. This represents a novel task since the majority of the literature focused on Minkowskian frames or on homogeneous backgrounds. Let us start from the Cartesian intervals analyzed above. If we center our frame in spherical coordinates at $x_{\rm A}$, then we see that $\varDelta x = r$ provided the angular coordinates are $\theta_{\rm A} = \theta_{\rm B}$ and $\phi_{\rm A} = \phi_{\rm B}$. Thus, we can rewrite Eq.\ \eqref{quanpr} as 
\begin{equation}
\label{quanradpres}
q(r) = \left|r \pm  \frac{\ell_*}{\alpha}\left(\frac{r}{\ell_*}\right)^\alpha\right|\,.
\end{equation}
Here we have defined $q(r) \equiv \varDelta q(r)$. In the deterministic view, this formula states that the radius acquires a non-linear modification whose sign depends on the presentation. In the stochastic view, we do not have any non-linear correction of the radius but, rather, the latter is afflicted by an intrinsic stochastic uncertainty and it fluctuates randomly between $r + \frac{\ell_*}{\alpha}({r}/{\ell_*})^\alpha$ and $r - \frac{\ell_*}{\alpha}({r}/{\ell_*})^\alpha$. In the first case, we just have a deformation of the radius, while in the second case we are suggesting that a stochastic (most likely quantum \cite{minlen}) feature comes out as a consequence of multi-fractional effects, namely the radius acquires a sort of fuzziness due to multi-fractional effects. 

Including also one mode of log oscillations, which are present in the most general multi-fractional measure \cite{first}, in the spherical-coordinates approximation Eq.\ \eqref{quanradpres} is modified by a modulation term:
\bs\label{logmeas}\ba
q(r) &=& \left|r \pm  \frac{\ell_*}{\alpha}\left(\frac{r}{\ell_*}\right)^\alpha F_\omega(r)\right|,\\
F_\omega(r) &=&1+A\cos\left(\omega\ln\frac{r}{\ell_\infty}\right)+B\sin\left(\omega\ln\frac{r}{\ell_\infty}\right)\!.
\ea\es
Here $A<1$ and $B<1$ are arbitrary constants and $\omega$ is the frequency of the log oscillations. The ultra-microscopic scale $\ell_\infty$ is no greater than $\ell_*$ and can be as small as the Planck length \cite{rev,minlen}. Notice that the plus sign is for the initial-point presentation, the minus for the final-point one, and both signs are retained in the interpretation of the multi-fractional modifications as stochastic uncertainties. The polynomial part of Eq.\ \Eq{logmeas} features the characteristic scale $\ell_*$ marking the transition between the ultraviolet and the infrared, regimes with a different scaling of the dimensions. On the other hand, the oscillatory part $F_\omega(r)$ is a signal of discreteness at very short distances, due to the fact that it enjoys the discrete scale invariance $F_\omega(\lambda_\omega r)=F_\omega(r)$, where $\lambda_\omega=\exp(-2\pi/\omega)$. Averaging over log oscillations yields $\langle F_\om\rangle =1$ and Eq.~\Eq{quanradpres} \cite{frc2}.  Indeed, in the stochastic view, the logarithmic oscillatory part is  regarded as the distribution probability of the measure that reflects a non-trivial microscopic structure of fractional spaces \cite{minlen}. 

We want to take expression \Eq{quanradpres} or the more general \Eq{logmeas} as our definition of the radial geometric coordinates, while we leave the measure trivial along the remaining $2+1$ directions $(t,\theta,\phi)$. We will consider modifications in the radial and/or time part of the measure for the theory with weighted derivatives, while still leaving the angular directions undeformed. Note that $q(r) \neq \sqrt{[q^1(x^1)]^2+[q^2(x^2)]^2+[q^3(x^3)]^2}$ (assuming the spherical system is centered at $x^\mu=0$). In fact, we derived Eq.\ \eqref{quanradpres} passing to spherical coordinates in the fractional frame and, of course, this is not equivalent to having geometric spherical coordinates \cite{trtls} as in \Eq{quanradpres}. However, it is not difficult to convince oneself that the difference between $q(r)$ and $\sqrt{[q^1(x^1)]^2+[q^2(x^2)]^2+[q^3(x^3)]^2}$ is negligible with respect to the correction term in \Eq{quanradpres} at sufficiently large scales, which justifies the use of the spherical geometric coordinate $q(r)$ as a useful approximation to the problem at hand. Notice, incidentally, that the geometric radius in the theory with fractional derivatives is $q(r)$ exactly \cite{rev}.

In the Sect.\ \ref{IIBa}, we will analyze the effects of this multi-fractional radial measure on black-hole horizons and, consequently, on the Hawking temperature of evaporation. We will see that, as announced, the initial-point presentation and the final-point presentation will produce different predictions both from a qualitative and a quantitative point of view. For instance, in the absence of log oscillations, according to the initial-point presentation we shall find the horizon radius $r_{\rm h} \simeq r_0 - \delta$, where $r_0=2MG$ is the usual Schwarzschild radius and $\delta>0$ will be introduced later (i.e., a smaller horizon with respect to GR), while $r_{\rm h} \simeq r_0 + \delta$ (a bigger horizon with respect to GR) if we choose the final-point presentation. On the other hand, in the stochastic view the results obtained with the initial-point and the final-point presentations will be interpreted as the extremes of fluctuations of relevant physical quantities. Then the horizon will be the one of GR but it will quantum-mechanically fluctuate around its classical value, $r \simeq r_0 \pm \delta$. This shows how non-trivial local quantum-gravity features can modify macroscopic properties such as the structure of black-hole horizons. 

To summarize, we are going to analyze the multi-fractional Schwarzschild solution in six different cases:
\begin{enumerate}
\item in the deterministic view with the initial-point presentation;
\item in the deterministic view with the final-point presentation;
\item in the stochastic view, where the presentation ambiguity corresponds to an intrinsic uncertainty on the length of the fractional radius,
\end{enumerate}
without and with log oscillations.


\subsection{Horizons}\label{IIBa}

Looking at Eqs.\ \eqref{riemten}--\eqref{einact} and recalling the related discussion, it is easy to realize that the Schwarzschild solution in geometric coordinates $q^\mu(x^\mu)$ (as well as all the other GR solutions) is a solution of the $q$-multi-fractional Einstein equations. Explicitly, the Schwarzschild line element in the multi-fractional theory with $q$-derivatives is given by
\ba
\prescript{q}{}{\rmd s^2} &=& -\left[1-\frac{r_0}{q(r)}\right] \rmd t^2 + \left[1-\frac{r_0}{q(r)}\right]^{-1}\rmd q^2(r)\nonumber\\
&&+ q^2(r)(\rmd\theta^2 + \sin^2\theta \rmd\phi^2),\label{qBH}
\ea
where $r_0:=2GM$, $M$ is the mass of the black hole and $q(r)$ is a non-linear function of the radial fractional coordinate $r$, given by Eq.\ \eqref{quanradpres} in the case of the binomial measure without log oscillations and by Eq.\ \eqref{logmeas} in their presence. 

Our first task is to study the position of the event horizon. As anticipated, fixing the presentation we will find that the horizon is shifted with respect to the standard Schwarzschild radius $r_0$. In particular, choosing the initial-point presentation the radius becomes smaller, while it is larger than the standard value $2GM$ in the case of the final-point presentation. The two shifted horizons obtained by fixing the presentation can also be regarded as the extreme fluctuations of the Schwarzschild radius, if we interpret the presentation ambiguity as an intrinsic uncertainty on lengths coming from a stochastic structure at very short distances (or, equivalently, as a semi-classical quantum-gravity effect) according to \cite{minlen}. From this perspective, the horizon remains $r_0$ but now it is affected by small quantum fluctuations that become relevant for microscopic black holes with masses close to the multi-fractional characteristic energy $E_*\propto 1/\ell_*$, i.e., when the Schwarzschild radius becomes comparable with the multi-fractional correction. 

\begin{figure}[ht!]
\centering
\includegraphics[width=3in]{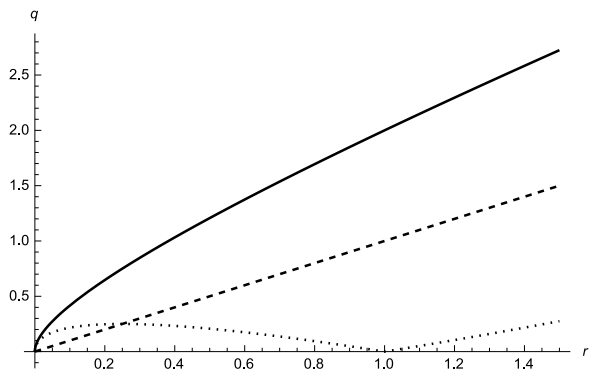}
\caption{Behavior of the geometric radius $q(r)$ as a function of the fractional radius $r$ with $\alpha = 1/2$ and $\ell_* =1$. The \textit{solid line} is the relation for the initial-point representation; choosing instead the final-point presentation, we find the \textit{dotted line}; the \textit{dashed line} is the ordinary case $q(r)=r$. The interpretation of multi-fractional corrections as quantum/stochastic uncertainties would make the \textit{dashed line} fuzzy by adding random fluctuations between the two other curves in the limit of large fractional radius $r$. Then, once we enter into the regime where $r \sim \delta r$ (i.e., $r<1$ in the plot), it is no longer allowed to talk about a radial distance $r$ according to the stochastic view.}
\label{rad}
\end{figure}

From Eq.\ \eqref{qBH}, the equation that determines the fractional event horizon $r_{\rm h}$ is
\be
q(r_{\rm h})= r_0\,,
\ee
valid even for the most general multi-fractional measure (which we have not written here but can be found in \cite{first}). Looking at this implicit formula for $r_{\rm h}$ in the case \Eq{logmeas}, it is evident that the initial-point $r_{\rm h}$ is inside the Schwarzschild horizon and, on the opposite, the final-point $r_{\rm h}$ stays outside the Schwarzschild horizon. However, in order to make an explicit example and also to get quantitative results, let us restrict ourselves to the coarse-grained case without log oscillations. Then the above equation simplifies to
\begin{equation}\label{horeq}
r_{\rm h} \pm \frac{\ell_*^{1-\alpha}}{\alpha}r_{\rm h}^\alpha = r_0.
\end{equation}
If we also fix the exponent by choosing $\alpha = 1/2$ (a value that has a special role in the theory \cite{rev,minlen}), we can easily solve the horizon equation analytically, obtaining
\begin{equation}\label{sol1}
r^{\rm ip}_{\rm h} = 2\ell_* + r_0 -2\sqrt{\ell_*^2 + r_0\ell_*}<r_0
\end{equation}
for the initial-point presentation, while
\begin{equation}\label{sol2}
r^{\rm fp}_{\rm h} =  r_0 +2\sqrt{\ell_*^2 + r_0\ell_*}- 2\ell_*>r_0
\end{equation}
for the final-point presentation. The superscripts distinguish the two possibilities. On the other hand, following the interpretation of \cite{minlen}, we would have
\begin{equation}
\label{fluct}
r_{\rm h} = r_0 \pm \delta(r)\,,\qquad \delta(r) := 2\sqrt{\ell_*^2 + r_0\ell_*}- 2\ell_*\,,
\end{equation}
where $\delta$ has the meaning of uncertainty on the position of the event horizon generated by the intrinsic stochasticity of spacetime. In Fig.\ \ref{rad}, we show the geometric radius $q(r)$ as a function of the fractional radius $r$ for the two different presentations we consider.


\subsection{Singularity}\label{IIBb}

The next task is to study whether and how the curvature singularity of the Schwarzschild solution is affected by multi-fractional effects. The bottom line is that the singularity is still present but the causal structure of black holes generally changes. In fact, novel features appear both for the final-point presentation and the case of a fuzzy radius. Consider first the measure without logarithmic oscillations. (i) In both the initial-point and the final-point presentations, there is no departure from the GR prediction on the curvature singularity at the center of the black hole, since $q(0)=0$ (for the most general factorizable measure). (ii) However, and contrary to what one might have expected, if we choose the final-point presentation, a second essential singularity appears. In fact, the geometric radius in the final-point presentation $q(r)=r-({\ell_*^{1-\alpha}}/{\alpha})r^\alpha$ has two zeros where the line element \eqref{qBH} diverges, one at $r=0$ and one at the finite radius
\begin{equation}\label{r2}
r = \alpha^{-\frac{1}{1-\alpha}}\ell_*\quad\Rightarrow\quad r\sim\ell_*\,.
\end{equation}
The second expression stems from the fact that the second flow-equation theorem \cite{first} leaves freedom in picking the prefactor ``$\ell_*/\a$'' in Eq.\ \Eq{mea} and, making an $\a$-independent choice, the numerical factor in \Eq{r2} is always $O(1)$. This $r=\ell_*$ locus corresponds to a ring singularity that is not present in the Schwarzschild solution of GR. (iii) Finally, in the stochastic view the reader might guess that the singularity is resolved due to multi-fractional (quantum) fluctuations of the measure. Unfortunately, this is not the case. In fact, the origin $r= 0$ represents a special point because $\delta(0)=0$ and it does not quantum fluctuate. Therefore, in the origin multi-fractional effects disappear and the theory inherits the singularity problem of standard GR. Let us also mention that stochastic fluctuations become constant in the limit $\alpha \rightarrow 0$ and the singularity might actually be avoided. However, $\a=0$ is not a viable choice in the parameter space, unless log oscillations are turned on. We will do just that now.

\begin{figure}[ht!]
\centering
\includegraphics[width=3in]{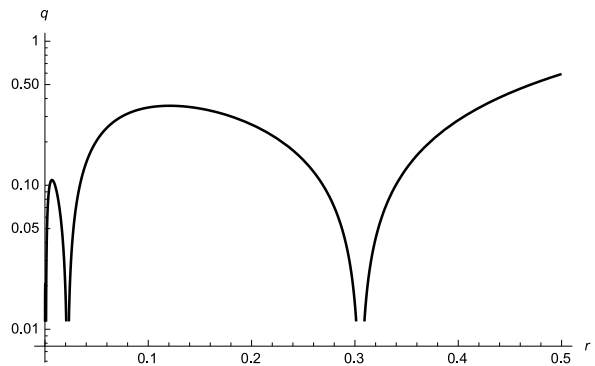}
\caption{Semi-log graph showing the behavior of the geometric radius \eqref{logmeas} as a function of the fractional radius $r$, with $A = B = \omega = 1$. In this figure, $\ell_* = 1$ and $\ell_\infty =10^{-2}$. Here $\alpha = 1/2$ and we chose the initial-point presentation. There are periodic zeros of $q(r)$ which are additional singularities for $x \ll \ell_*$, at ultra-short distances from the origin. The qualitative trend does not depend on the chosen presentation. Cosmological observations constrain the amplitudes in the measure to values that avoid these singularities.}
\label{rad2}
\end{figure}

Considering the full measure \eqref{logmeas}, we find that not only is the singularity not resolved, but in principle there may also be other singularities for $r \neq 0$
due to discrete scale invariance of the modulation factor $F_\omega(r)$. To see this in an analytic form, we first consider a slightly different version of the log-oscillating measure \Eq{logmeas}, $q(r)=[r+(\ell_*/\a)(r/\ell_*)^\a]F_\omega(r)$, where the modulation factor multiplies also the linear term. This profile is shown in Fig.\ \ref{rad2}. The geometric radius vanishes periodically at $r=\exp(-n\b_\pm)\ell_\infty$, $n=0,1,2,\dots$, where $\b_\pm={\rm arccos}[(A\pm B \sqrt{A^2+B^2-1})/(A^2+B^2)]$. Since $-1\leq A\leq 1$ and $-1\leq B\leq 1$, the parameter $\b_\pm$ is well defined only when $|B|\geq \sqrt{1-A^2}$. In general, also in the actual case \Eq{logmeas}, these extra singularities appear only when one or both amplitudes $A$ and $B$ take the maximal value $|A|\sim 1\sim|B|$. Fortunately, observations of the cosmic microwave background constrain the amplitudes to be smaller than about $0.5$ \cite{Calcagni:2016ofu}, which means that some protection mechanism avoiding large log oscillations is in action. This is also consistent with the fact that, in fractal geometry, these oscillations are always tiny ripples around the zero mode.


\subsection{Evaporation temperature}\label{IIBc}

\begin{figure}[ht!]
\centering
\includegraphics[width=3in]{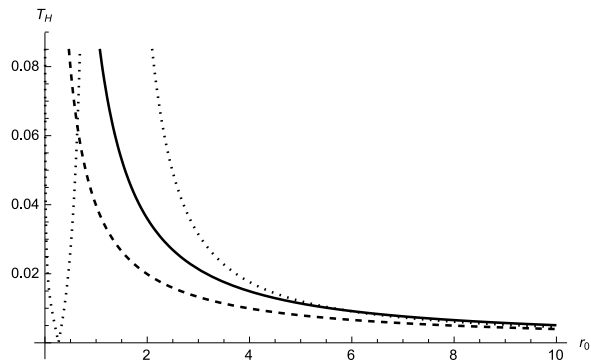}
\caption{\label{TH}Behavior of the Hawking temperature $T_{\rm H}$ as a function of  $r_0$, for $\alpha = 1/2$ and $\ell_* = 1$. The \textit{solid line} is the black-hole solution in the multi-fractional theory with $q$-derivatives in the initial-point presentation; the \textit{dotted line} is for the final-point presentation; the \textit{dashed line} represents $T_{\rm H}$ for the GR Schwarzschild solution. If we regard the multi-fractional part as a quantum uncertainty on the radius and the presentation ambiguity as the two possible signs for the fluctuations, then $T_{\rm H}$ would quantum fluctuate around the standard Hawking temperature.}
\end{figure}

We continue the analysis of the Schwarzschild solution in multi-fractional gravity with $q$-derivatives by studying the thermodynamics of the black hole in the absence of log oscillations. In particular, we calculate the Hawking temperature for both presentations and compare it with the GR case.  In the presence of logarithmic oscillations of the measure, the Hawking temperature collapses to the standard behavior in the limit of large $r_0$, while for small radii we encounter a series of poles in correspondence with the zeros of the geometric radius (see the previous subsection and, in particular, Fig.\ \ref{rad2}). The Hawking temperature can be defined in the following manner: 
\begin{equation} \label{eq:thq}
T^{\rm ip,fp}_{\rm h} := \frac{1}{4\pi}\frac{\rmd}{\rmd r}\left[1-\frac{r_0}{q(r)}\right]\Bigg\vert_{r = r^{\rm ip,fp}_{\rm h}}. 
\end{equation}
Imposing the same restrictions we made above for the horizon, we can find the analytic expression for the multi-fractional Hawking temperature:
\begin{eqnarray}
&&T^{\rm ip}_{\rm h} =\frac{r_0\left(1+ \sqrt{\frac{\ell_*}{2 r^{\rm ip}_{\rm h}}} \right)}{4\pi \left(r^{\rm ip}_{\rm h} + \sqrt{2 \ell_* r^{\rm ip}_{\rm h}}\right)^2},\label{thipfp1} \\
&&T^{\rm fp}_{\rm h} =\frac{r_0\left|1- \sqrt{\frac{\ell_*}{2 r^{\rm fp}_{\rm h}}} \right|}{4\pi \left(r^{\rm fp}_{\rm h} - \sqrt{2 \ell_* r^{\rm fp}_{\rm h}}\right)^2},\label{thipfp2}
\end{eqnarray}
which, of course, reduce to $ \lim_{\ell_* \rightarrow 0}T^{\rm ip,fp}_{\rm h} = T_{{\rm H}0}:= 1/(4\pi r_0)$ in the standard case. As expected, there are no appreciable effects at large distances $r_0 \gg \ell_*$ and the correct GR limit is naturally recovered. Given that, we can ask ourselves what happens to micro (primordial) black holes with Schwarzschild radius close to or even smaller than $\ell_*$. Again we shall discuss all the three possibilities regarding the presentation. Let us start with the initial-point case and make an expansion of Eq.~(\ref{thipfp1}) up to the first order in $\ell_*$ for $r_0\ll \ell_*$:
\begin{equation}\label{smallqT}
T_{\rm H} \simeq \frac{\ell_*}{2\pi r_0^2} = \frac{2\ell_*}{r_0} T_{{\rm H}0} > T_{{\rm H}0}\,.
\end{equation}
Thus, multi-fractional micro black holes are hotter than their GR counterparts, which means that they should also evaporate more rapidly. Such a result is somehow counter-intuitive since we found that, in presence of putative quantum-gravity effects (here consisting in a non-trivial measure), not only is the information paradox \cite{infopar1,infopar2,infopar3,infopar4,infopar5} not solved, but it even gets worse.\footnote{See Ref.~\cite{infopar6} for a similar conclusion in the context of loop-quantum-gravity black holes.} This can be noticed immediately by comparing the solid line in Fig.\ \ref{TH} with the usual behavior represented by the dashed line.

In the final-point presentation, the modification of the Hawking temperature is given by Eq.\ \eqref{thipfp2}, where the event horizon at which $T^{\rm fp}_{\rm h}$ has to be evaluated is defined in Eq.\ \eqref{sol2}. As the reader can easily understand by looking at Fig.\ \ref{TH}, the behavior is even worse with respect to the initial-point case. In fact, the dotted line (which represents $T^{\rm fp}_{\rm h}$ as a function of $r_0$) increases more rapidly than the other two curves as the black-hole mass decreases. Therefore, again we find that multi-fractional effects do not cure the GR information paradox but make it even more prominent. However, it is interesting to look at the behavior of $T^{\rm fp}_{\rm h}$ for very small black holes. We can see that there is a value of $r_0$ where the Hawking temperature vanishes. Thus, in multi-fractional $q$-gravity in the final-point presentation, (micro) black holes with $r_0 = (5-2 \sqrt{5})\ell_* \approx 0.5 \ell_*$ do not emit Hawking radiation. Even so, however, they are unstable since, as clear from the figure, any increase $+\delta M$ or decrease $-\delta M$ of their mass would make them emitting rather efficiently. 

The third possibility is to regard multi-fractional modifications as an uncertainty on relevant physical quantities. In that case, we have $T_{\rm H} = T_{{\rm H}0} \pm \delta T $, i.e., the Hawking temperature fluctuates around the GR value. As for the other quantities we analyzed, the magnitude of such random fluctuations depends on how large $\ell_*$ is and it decreases as $M$ (or, equivalently, $r_{\rm h}$) increases. 

To summarize, the theory with $q$-derivative does not solve the information paradox of GR, a datum consistent with the problems one has when quantizing gravity perturbatively here \cite{rev}. On the other hand, approximating the theory to the stochastic view the information paradox is not worsened and the role of the random fluctuations in this respect is not yet clear. This may indicate that the theory with fractional derivatives is better behaved than its approximation the $q$-theory, again consistent with previous findings \cite{rev}.


\subsection{Effects on the quantum ergosphere}\label{IIC}

In Ref.\ \cite{CalcagniArzano}, it was shown that the recent discovery of gravitational waves can provide, at least in principle, a tool to place observational constraints on non-classical geometries. In particular, a way to obtain an upper bound on the multi-fractional length $\ell_*$ consists in comparing the mass shift $\varDelta M$, due to quantum fluctuations of the horizon, with the experimental uncertainty $\delta M_{\rm BH}$ on the mass of the final black hole in the GW150914 merger. Such a mass shift $\varDelta M$ can be related to the appearance of a quantum ergosphere (see Refs.\ \cite{CalcagniArzano,Arzano}). Here we want to reconsider this analysis in the framework of the multi-fractional theory with $q$-derivatives. In other words, we are going to study the formation of the quantum ergosphere in the multi-fractional Schwarzschild black hole \eqref{qBH} with the objective to see if it is possible to find constraints on $\ell_*$. In this section only, we ignore log oscillations.

The mass shift $\varDelta M$ is related to a corresponding change of the radial hypersurfaces $\varDelta q(r)$ by $\varDelta q(r)=2\varDelta M\, G$. In order to find the width $\varDelta r$ of the ergosphere, we have to plug Eq.\ \eqref{quanradpres} into the above expression, thereby obtaining 
\begin{equation}
\label{deltaM}
\frac{\varDelta r}{2} = \frac{\varDelta M\, G}{1\pm({\ell_*}/{r})^{1-\alpha}} =: \varDelta \widetilde{M}\, G\,,
\end{equation}  
where the plus sign holds for the initial-point presentation and the minus sign for the final-point presentation. According to Ref.\ \cite{CalcagniArzano}, noting that $\varDelta M \sim E^2_*/M_{\rm BH}$ in the absence of multi-fractional effects \cite{Arzano} (here $E_*$ is some quantum-gravity scale) and imposing $\varDelta M < \delta M_{\rm BH}$ (with $\delta M_{\rm BH}=O(M_{\odot}$) if we are considering the GW150914 merger), one obtains a very high bound, $E_* < 10^{58}\,{\rm GeV}$. In the case of the multi-fractional theory with $q$-derivatives, if we use the initial-point presentation then we have a plus sign in the denominator of Eq.\ \eqref{deltaM} and the energy bound is even higher, since $\varDelta \widetilde{M} < \varDelta M$. Things completely change with the final-point presentation, where the upper bound on $E_*$ is
\ba
E_* &<& \sqrt{M_{\rm BH} \delta M_{\rm BH}\left[1-\left(\frac{\ell_*}{r}\right)^{1-\alpha}\right]}\nonumber\\
&=&\sqrt{1-\left(\frac{\ell_*}{r}\right)^{1-\alpha}}10^{58}{\rm GeV}\,.
\ea
For $\alpha \ll 1$, the upper bound remains $E_* < 10^{\rm 58}\,{\rm GeV}$ for any sensible value of $\ell_*$. However, in the limit $\alpha \rightarrow 1$ the upper bound dramatically lowers, regardless how small is the ratio $\ell_*/r$. This shows that the correction to the quantum ergosphere, combined with gravitational waves measurements, can be used to severely constrain the multi-fractional theory with $q$-derivatives in the final-point presentation for big values (i.e., close to $1$) of $\alpha$. Note, however, that values $\alpha\sim 1$ do not have any theoretical justification.

Adopting the stochastic view instead, the correction term in the denominator of Eq.\ \eqref{deltaM} would result from the quantum uncertainty on the radius, i.e., $q(r) = r \pm \delta r$. Given that, the only constraint coming from the quantum-ergosphere calculation is  $({\ell_*}/{r})^{1-\alpha}<1$. However, this inequality is always satisfied as far as we consider solar-mass or super-massive black holes for which the radius $r$ of the ergosphere exceeds the multi-fractional length $\ell_*$ by several orders of magnitude. In this case, multi-fractional effects on the ergosphere might be relevant only for primordial (microscopic) black holes with $r<\ell_*$. On the other hand, according to the stochastic view, it is meaningless to contemplate distances smaller than the multi-fractional uncertainty $\delta r$. Consequently, we conclude that this argument cannot be used to constrain the scale $\ell_*$.


\section{Schwarzschild solution in the multi-fractional theory with weighted derivatives}\label{III}

The gravitational action in the theory with weighted derivatives is similar to the one of scalar-tensor models, with the crucial difference that the role of the scalar field is played by the non-dynamical measure weight $v(x)=v_0(x^0)\ldots v_{D-1}(x^{D-1})$, where $v_\mu(x^\mu)=\p_\mu q^\mu(x^\mu)$. Since this is a fixed profile in the coordinates, one does not vary the action with respect to it and the dynamical equations of motion are therefore different from the scalar-tensor case. However, even if it is not dynamical, the measure profile affects the dynamics of the metric so much that the resulting cosmologies depart from the scalar-tensor case \cite{frc11}.

As for scalar-tensor models, we can identify a ``Jordan frame'' (or fractional picture) and an ``Einstein frame'' (or integer picture) related to each other by a measure-dependent conformal transformation of the metric. In the Jordan frame, the action for multi-fractional gravity with weighted derivatives in the absence of matter is given by \cite{frc11}
\be
S_g = \frac{1}{2\kappa^2} \int d^D x e^{\Phi/\beta} \sqrt{-g} \left[R-\Omega \partial_\mu \Phi\partial^\mu\Phi -U(v) \right],\label{eq:acjor}
\ee
with 
\be 
\Omega :=  \frac{9\omega}{4\beta^2}e^{\frac{2}{\beta}\Phi} + (D-1) \left(\frac{1}{2\beta_*}-\frac{1}{\beta}\right),
\ee
where $\Phi(x) = \ln v(x)$ is \emph{not} a Lorentz scalar field and $\omega$ is an arbitrary constant (not to be confused with the frequency of log oscillations). In $D=4$ topological dimensions, $\beta = \beta_* = 1$ is fixed by the theory. In \cite{frc11}, one demanded that $U\neq 0$ in order to support consistent solutions with cosmological constant. Since this quantity is measure-dependent but background independent, if we want to describe both black holes and consistent cosmologies, we have freedom to choose $\Omega$ but not $U(v)$. However, keeping black holes and cosmology as separate entities this restriction is lifted.

The metric $g_{\mu\nu}$ in the Jordan frame is not covariantly conserved, just like in a Weyl-integrable spacetime. For convenience, we will move to the Einstein frame, which is obtained after performing the Weyl mapping
\be \label{eq:weylmap}
g_{\mu\nu} = e^{-\Phi}\overline{g}_{\mu\nu}, 
\ee
so that the action \Eq{eq:acjor} in $D=4$ reads
\be\label{Sgei}
S_g = \frac{1}{2\kappa}\int d^4 x \sqrt{-\overline{g}} \left(\overline{R}-\Omega\partial_\mu \Phi \overline{\partial}^\mu \Phi - e^{-\Phi} U \right).
\ee

In this frame, although the metricity condition $\overline{\nabla}_\sigma g_{\mu\nu}=0$ is satisfied, the dependence in the measure profile cannot be completely absorbed.  As we will see, black-hole solutions are highly sensitive to the choice of $\omega$, which may even hinder their formation. For illustrative purposes, we will examine the cases $\Omega=0$ ($\omega$ fixed) and $\Omega=-3/2$ ($\omega=0$). At this point, it is important to recall a key feature of these theories. In standard GR, at the classical level one has the freedom to pick either the Jordan or the Einstein frame, leading to equivalent predictions; at the quantum level, these frames are inequivalent and one must make a choice based on some physical principle. In the multi-fractional case, the existence of the non-trivial measure profile $v(x)$ that modifies the dynamics renders both frames physically inequivalent already at the classical level. A natural question is which one is ``preferred'' for observations. The answer is the following. Measurements involve both an observable and an observer. Given the nature of multi-scale spacetimes, both feel the anomalous geometry in the same way if they are characterized by the same scale, while they are differently affected by the geometry otherwise. This is due to the fact that measurement apparatus have a fixed scale and do not adapt with the changing geometry. In the multi-fractional field theory with weighted derivatives and in the absence of gravity, this occurs in the fractional picture, while in the integer picture the dynamics reduces to that of an ordinary field theory. In the presence of gravity, the integer picture (Einstein frame) is no longer trivial (see Eq.\ \Eq{Sgei}), but the interpretation of the frames remains the same. Therefore, the Jordan frame is the physical one \cite{frc11}. Physical black holes as those found in astrophysical observations can be formally described within the Einstein frame, while to extract observables one has to move to the Jordan frame.


\subsection{Black-hole solution with \texorpdfstring{$\Omega=0$}{}}\label{IIIA} 

In this section, we will examine the spherically symmetric solution when the ``kinetic term'' of the measure vanishes:
\begin{eqnarray}
\Omega &=& 0\quad \Rightarrow\quad \omega = \frac{2}{3}e^{-2\Phi},\\
S_g &=& \frac{1}{2\kappa^2}\int d^4 x  \sqrt{-\overline{g}} \left(\overline{R} -e^{-\Phi} U \right).
\end{eqnarray}
Taking the variation with respect to $\overline{g}_{\mu\nu}$,\footnote{Since $\Phi$ is a not dynamical measure profile, we do not vary the action \Eq{eq:acjor} with respect to it.} we get
\be 
\overline{R}_{\mu\nu} -\frac{1}{2}\overline{g}_{\mu\nu}\left(\overline{R} -e^{-\Phi} U\right) = 0.
\ee
We restrict to an isotropic, static and radially symmetric geometry. Thus, our \emph{Ansatz} is
\ba
\overline{g}_{\mu\nu}\rmd x^\mu \rmd x^\nu &=& -\g_1(r)\rmd t^2 + \g_2(r) \rmd r^2\nonumber\\
&& +\g_3(r) r^2\left(\rmd\theta^2 + \sin^2\theta\, \rmd\phi^2 \right).\label{gbh}
\ea
After some manipulations ($\g_3$ can be consistently set to 1), the Einstein equations read (primes denote derivatives with respect to $r$ and the $r$ dependence is implicit in all functions)
\bea
0 &=& (\g_1 \g_2)', \\
0 &=& \g_1''-\g_1' \left(\frac{\g_2'}{2 \g_2}+\frac{1}{r}-\frac{\g_1'}{2 \g_1}\right)-\frac{\g_1}{r^2}\left(r \frac{\g_2'}{\g_2} -2 \g_2 +2 \right),\nonumber\\
\eea
plus a master equation for $U$:
\be \label{eq:u}
U = -v\frac{2}{\g_2r} \left(\frac{\g_1'}{\g_1}-\frac{1-\g_2}{r}\right).
\ee

Restoring coordinate dependence, a consistent solution is given by
\be\label{eq:sola}
\g_1(r) = 1-\frac{r_0}{r} \pm \frac{\chi}{6} r^2, \quad \g_2(r) = \frac{1}{\g_1(r)}, \quad U = \mp v(x)\chi,
\ee
which is a two-parameter family with a cosmological potential. Several caveats are in order. First, although the functions $\g_1$ and $\g_2$ depend only on the radius, the ``potential'' term $U$ is factorized in the coordinates, since it depends on the measure weight $v(x)$ (which we did \emph{not} approximate by a radial profile as done in the theory with $q$-derivatives). Second, the existence of the ``hair'' $\chi={\rm const}$ was foreseeable since we have considered a non-zero ``potential'' coupled to gravity. Third, the sign in front of the $r^2$ term is arbitrary but, in order to get a Schwarzschild--de Sitter solution, we pick the minus sign. In \cite{chi1,Kagramanova:2006ax}, the cosmological constant $\chi$ was expressed in terms of a temperature $T_{\rm vac}$ by means of the Stefan--Boltzmann law, so that
\be \label{eq:chivalue}
\chi = \frac{4\pi^3}{15}\frac{T_{\rm vac}^4}{m_{\rm Pl}^2} \approx 10^{-66} \, \text{eV}^{2},
\ee
where $T_{\rm vac}\approx 34\, \rm K$ \cite{Adams}.\footnote{We have employed the conversion factor $1\, \text{K} = 8.6217\times 10^{-5}\, \text{eV}$ and $m_\Pl = 1.22\times 10^{28}\, \text{eV}$.} However, in this scenario, the Stefan--Boltzmann law receives a sub-leading contribution as a consequence of integrating out in the presence of some measure profile, i.e., $\int^\infty_0 \rmd\nu \to \int^\infty_0 \rmd\nu\,  w(\nu),$ with $w(\nu) = 1 + \delta w(\nu)$, so that $\rho =\sigma T^4 + \delta \rho$ (for more details, see \cite{Calcagni:2016ofu}). Nevertheless, the correction is small. Taking for instance the binomial measure
\be 
\omega(\nu) = \left(1 + \Bigg\vert{\frac{\nu}{\nu_*}}\Bigg\vert^{1-\alpha}\right)^3, 
\ee
with $\alpha = 1/2, \quad \nu_* \simeq 3\times 10^{9}\, \rm{cm}^{-1}$ \cite{Calcagni:2016ofu}, and integrating out over all frequencies,
\ba
\rho &=& 240 \sigma \int^\infty_0 \rmd\nu \, w(\nu) \frac{\nu^3}{e^{\frac{2\pi \nu}{T}}-1} \sim \sigma T^4 + \delta \rho\,,\\
\delta \rho &=&  \sigma T^4 \left[\frac{4725}{16 \sqrt{2} \pi ^4} \sqrt{\frac{T}{\nu_*}} \zeta \left(\frac{9}{2}\right) \right] + \mathcal{O}\left( \frac{T}{\nu_*}\right), 
\ea
we get $\delta \rho(T_{\rm vac})/(\sigma T_{\rm vac}^4) \approx 10^{-4}$, which becomes even smaller for lower temperatures. Since we are interested only in the order of magnitude of $\chi$, we can just adopt the standard power law
\be \label{eq:stef}
\rho\sim T^4, 
\ee 
and set the value of $\chi$ as in \Eq{eq:chivalue}, ignoring any other anomalous contribution. Moreover, according to \Eq{eq:chivalue}, one sees that even for a black hole of mass $10^{10}M_\odot$, it is safe to assume that $(G M)^2\ll 1/\chi$.

Let us pause for a moment and discuss one of the main results of this paper. Just assuming a non-trivial dimensional flow in the Hausdorff dimension of spacetime (i.e., a non-trivial multi-fractional measure), we have just shown that the simplest black-hole solution is the Schwarzschild--de Sitter solution, where the cosmological constant term is caused by the multi-scaling nature of the geometry. This offers a possible reinterpretation of the cosmological constant \cite{cosmocon1,geon} as a purely geometric term arising from the scaling properties of the integration measure. Since, in this case, there is no reason to expect a huge value of $\chi$ due to quantum fluctuations of the vacuum energy (as it would be the case in quantum field theory), then we do not have the problem of fine tuning large quantum corrections. This step towards the solution of the cosmological constant problem is somehow analogous to what happens in unimodular gravity, as noted in \cite{frc11}. In unimodular gravity, as a consequence of fixing the determinant of the metric $g_{\mu\nu}$, the source of the gravitational field is given only by the traceless part of the stress-energy tensor and, thus, all potential energy is decoupled from gravity (see, e.g., Ref.\ \cite{Alv1}). In this way, $\chi$ appears as an integration constant rather than a parameter of the Lagrangian \cite{Smolin,Alv2}. However, unimodular gravity also has the feature of breaking time diffeomorphisms as recognized for the first time in Ref.\ \cite{DamNg}, whose consequences are still to be completely understood. The multi-fractional scenario has the advantage of formally preserving full diffeomorphism invariance \cite{CR}, although in this case the ``diffeomorphism'' transformations are deformed with respect to those of general relativity.

At this point, it is interesting to discuss the causal structure of our manifold. Imposing $\g_1(r_{\rm h})=0$, we distinguish three horizon radii ($r_0 = 2MG$)
\be
r_{\rm h}^{(1,2)} \sim  -\left(M G \pm \sqrt{\frac{6}{\chi}}\right),\quad 
r_{\rm h}^{(3)} \sim  r_0\left( 1 + \frac{2}{3}M^2 G^2 \chi\right).
\ee
$r_{\rm h}^{(1)}$ is unphysical since it is negative. In order for $r_{\rm h}^{(2)}$ to be physical, it should be $\sqrt{{6}/{\chi}}>M G$, which means that in the small-$\chi$ limit $r_{\rm h}^{(2)}$ is the cosmological horizon. $r_{\rm h}^{(3)}$ is the apparent inner horizon which reduces to the standard Schwarzschild radius when $\chi\to 0$. Hereafter, we shall consider only this horizon.

Undoing the Weyl mapping, the solution in the Jordan frame is 
\be \label{eq:weylmap2}
g_{\mu\nu} = \frac{1}{v(x)} \overline{g}_{\mu\nu} .
\ee

Moreover, since in the Jordan frame the Hawking temperature is given by (recall that $\g_1(\text{r}_\text{h})=0$), 
\begin{eqnarray}
T_{\rm H}(x) & =&  \frac{1}{4\pi} \lim_{r\to r_{\rm h}} \Bigg\vert \frac{\g_1(r)}{v(x)}\Bigg\vert '\nonumber\\
&=&  \frac{1}{4\pi} \lim_{r\to r_{\rm h}} \Bigg\vert \frac{\g_1'(r)v(x)-\g_1(r) v'(x)}{v(x)^2}\Bigg\vert \nonumber\\
& = & \frac{1}{4\pi} \lim_{r\to r_{\rm h}} \Bigg\vert \frac{\g_1'(r) }{v(x)}\Bigg\vert = \frac{1}{4\pi}   \Bigg \vert \frac{r_0}{r_{\rm h}^2} -\frac{\chi}{3}r_{\rm h} \Bigg \vert \lim_{r\to r_{\rm h}} \frac{1}{v(x)} \nonumber\\
&=& T_{\rm H}^{(0)} \lim_{r\to r_{\rm h}} \frac{1}{v(x)},\label{HawTemp}
\end{eqnarray}
with $T_{\rm H}^{(0)} = |{r_0}/{r_{\rm h}^2} -{\chi}r_{\rm h} /3|/(4\pi)$, the Hawking temperature in the Einstein frame, it is immediate to notice a shift due to the anomalous geometry. From previous work \cite{rev}, we can safely infer that the contribution from the anomalous geometry to observables is rather tiny at large scales. Hence, we write 
\be \label{eq:vapprox}
v(x) \simeq 1 + \delta v(x) + \mathcal{O}(\delta v^2),
\ee
so that
\be 
T_{\rm H}{}_{(\Omega = 0)} \sim T_{\rm H}^{(0)} + \delta T_{(\Omega = 0)}
\ee
with 
\be 
T^{(0)}_{\rm H} = \left(T^{(0)}_{\rm BH} - T^{(0)}_{\rm vac}\right), \quad \delta T_{(\Omega = 0)} = -T_{\rm H}^{(0)}  \lim_{r\to r_{\rm h}} \delta v
\ee
and
\be 
T^{(0)}_{\rm BH} = \frac{M G}{2\pi r_{\rm h}^2}\simeq T_{{\rm H}0}, \quad T^{(0)}_{\rm vac} = \chi \frac{r_{\rm h}}{12\pi}\sim \frac{M G \chi}{3\pi},
\ee
where we have approximated $r_{\rm h}\sim r_{\rm h}^{(3)}$. Since $1>\delta v(x)>0$, one expects to get a redshift. Two comments are in order. The first is that the temperature now depends on the spacetime coordinates through the non-trivial measure profile $v(x)$, and, as stated before, this implies that one can have a spacetime-dependent redshift. The second is that the temperature has two sources: one is the standard black-hole temperature $T^{(0)}_{\rm BH}$ and the other, $T^{(0)}_{\rm vac}$, comes from the de Sitter background, can be related to the effective temperature scale of the cosmological vacuum energy. The equilibrium point is achieved when
\be \label{eq:mc}
T^{(0)}_{\rm BH} = T^{(0)}_{\rm vac} \to M= M_C \simeq \frac{1}{2 G}\sqrt{\frac{3}{2\chi}}.
\ee

This condition would set a critical mass scale $M_C$ above which accretion takes place at a higher rate than evaporation. Plugging in the $\chi$ estimate \Eq{eq:chivalue} ($G \propto 1/M_\Pl^2$), $M_C \approx 10^{52} \, \text{kg} \approx 10^{23}M_\odot$. Even for the largest monster black hole ever discovered so far, with $M\approx 10^{10}M_\odot$ \cite{nat1}, accretion cannot compete with evaporation.


\subsubsection{Consequences on the evaporation time of black holes}\label{IIIB}

It is interesting to ask oneself whether the anomalous geometry can lead to significant differences on the evaporation time of black holes, such extremely massive objects, with masses at least comparable with the solar mass, will have small Hawking temperatures. In particular, for this case, ${\delta \rho(T_{\rm H})}/(\sigma T_{\rm H}^4) \approx 10^{-8}$, so that the approximation \Eq{eq:stef} is well justified also here. According to the standard Stefan--Boltzmann law, the power emitted by a perfect black body in repose ($E=M$) is
\be \label{eq:power}
P = \sigma A_{\rm h} T_{\rm H}^4 = -\frac{\rmd E}{\rmd t} = -\dot{M},
\ee
$\sigma$ being the Stefan--Boltzmann constant. Note that, in this theory, the horizon area $A_{\rm h}$ remains unchanged
\ba
A_{\rm h} &=& \int \rmd\theta \rmd\phi\,v(x) \sqrt{g_{\theta\theta}g_{\phi\phi}} \Big\vert_{r=r_{\rm h}} = \int \rmd\theta \rmd\phi\,r^2 \sin \theta \Big\vert_{r=r_{\rm h}}\nonumber\\
&=& 4\pi r_{\rm h}^2.
\ea
For a process involving some energy (mass) loss, we compute the time needed to jump from an initial energy $E_{\rm i}$ to a final energy $E_{\rm f}$. Inserting \Eq{HawTemp} into \Eq{eq:power},
\be \label{eq:evap1}
-\frac{v(x)^4}{A_{\rm h} T^{(0)}_H{}^4} \rmd E\, \Bigg\vert_{r\to r_{\rm h}} = \sigma \rmd t.
\ee
At this point, we will consider a toy-model geometry where only the time and radial directions are anomalous, $v(x) = v_0(t) v_1(r)$, so that
\be \label{eq:pow1}
-\int^{E_{\rm f}}_{E_{\rm i}} \frac{v_1(r_{\rm h})^4}{r_{\rm h}^2 \left(T_{\rm H}^{(0)}\right)^4} \rmd E  = 4\pi \sigma \int^{t_{\rm f}}_{t_{\rm i}} \Bigg\vert\frac{1}{v_0(t)}\Bigg\vert^4 \,\rmd t.
\ee

Under the approximation \Eq{eq:vapprox}, the right-hand side of \Eq{eq:pow1} can be rewritten as
\be 
4\pi \sigma \int^{t_{\rm f}}_{t_{\rm i}} \, \big\vert 1- 4\delta v_0(t)\big \vert \, \rmd t .
\ee

Adopting the deterministic view with the initial-point presentation in this last part of the analysis, we set the binomial measure without log oscillations for each anomalous direction,
\bea \label{eq:binmeas}
v_0(t) &= & 1+ \delta v_0(t), \quad \delta v_0(t) = \Bigg\vert\frac{t_*}{t}\Bigg\vert^{1-\alpha_0}, \nonumber\\
v_1(r) &=& 1+ \delta v_1(r), \quad \delta v_1(r) = \Bigg\vert\frac{\ell_*}{r}\Bigg\vert^{1-\alpha}. 
\ea
Taking $r_{\rm h} \sim r_{\rm h}^{(3)}$, from \Eq{eq:pow1} we get
\ba
& &\frac{256}{15} \pi ^3 \Big\lbrace 5 G^2 \left(E_{\rm f}^3-E_{\rm i}^3\right) + 12  G \sqrt{2 G \ell_*} \left(E_{\rm f}^{5/2}-E_{\rm i}^{5/2}\right)  \nonumber\\
&&\quad+ G^3 \chi\left[28 G \left(E_{\rm f}^5-E_{\rm i}^5\right) + 60 \sqrt{2 G \ell_*} \left(E_{\rm f}^{9/2}-E_{\rm i}^{9/2}\right) \right]\Big\rbrace \nonumber\\
&&= 4\pi \sigma \, \left(\varDelta t - 4 \frac{t_*}{\alpha_0}\Bigg\vert \frac{\varDelta t}{t_*}\Bigg\vert^{\alpha_0} \right) ,\label{eq:prediction1}
\ea
with $\varDelta t = t_{\rm f} -t_{\rm i}$. Considering a process where we jump from an initial state to a final state with zero energy, for example the evaporation of a black hole, we have $E_{\rm i} = M_0$ (the initial mass) and $E_{\rm f} = M_{\rm f}=0$. Then
\ba
&&\frac{256}{3}\pi^3  G^2 M_0^3 + \frac{7168}{15} \pi ^3 G^4 M_0^5 \chi \nonumber\\
&&\quad+\frac{256}{15} \pi ^3 G M_0^2 \sqrt{2 G  M_0 \ell_*} \left(12+60 G^2 M_0^2 \chi \right) \nonumber\\
&& \simeq 4\pi \sigma \, \left( \varDelta t - 4 \frac{t_*}{\alpha_0}\Bigg\vert \frac{\varDelta t}{t_*}\Bigg\vert^{\alpha_0} \right).
\ea
Given some test black hole of mass $M_0 \approx M_{\odot}$, for the natural choice $\alpha_{0}= \alpha=1/2$ \cite{rev} and taking the most stringent characteristic time derived from $\alpha_{\rm QED}$ measurements \cite{frc13}, $t_* \approx 10^{-36} \, \rm s$, $\ell_* \approx 10^{-27}\, \rm m$, we get
\be \label{eq:timepre}
\Bigg\vert \frac{\left( \varDelta t\right)_0 -\varDelta t}{\varDelta t}\Bigg\vert \approx 10^{-16},
\ee
where we have employed Eq.\ \Eq{eq:chivalue} and $\left(\varDelta t\right)_0$ refers to the evaporation time predicted by the standard lore. Such deviation is independent of the presentation adopted. As it stands, multi-fractional effects entail slight changes on the evaporation time on black holes, therefore coinciding with the usual model in the large-scale regime.


\subsection{Black-hole solution with \texorpdfstring{$\Omega=-3/2$}{}}\label{IIIC}

The simplest version of multi-fractional gravity with weighted derivatives is in the absence of the fake ``kinetic'' term in the Jordan frame action, $\omega=0$ ($\Omega=-3/2$). In this case, the $v$ dependence cannot be eliminated in the equations of motion as we did before. The metric components now receive a direct contribution from the anomalous geometry, so that, in order to preserve staticity and radial symmetry, we have to consider a radial measure weight independent of angular coordinates, $v(x)=v(r)$.
This must be regarded as an approximation of the full theory because we do not have the liberty to change coordinates via a Lorentz transformation, which is not a symmetry of the theory.\footnote{On the other hand, the Fourier transform is well defined even when the measure weight is $v(r)$, as is clear from an inspection of the plane waves \cite{rev,frc3}.} As in the case with $q$-derivatives, the difference with respect to the exact case will be in sub-leading terms that do not change the qualitative features of the solution. Two other assumptions we will have to enforce in order to get an idea of the solution will be that of small geometric corrections and $\a =1/2$.
Having thus cautioned the reader, we can proceed. 

Considering a large-scale regime where multi-scale effects are small, $v_1(r) \simeq 1 + \delta v_1(r)$, for the black-hole metric \Eq{gbh} we have $\g_2=1/\g_1$ and
\be \label{eq:aexpansion}
\g_1 \simeq \tilde{\g_1} +  \delta \g_1 ,  \quad  \g_3 \simeq 1 +  \delta \g_3,
\ee
where $\tilde{\g_1}= 1-{r_0}/{r} -{\chi}r^2/{6}$. At zeroth order in the $M^2\chi $ expansion, the linearized Einstein equations are
\ba
0 &=& \frac{3 r }{2( r_0 - r) } \delta v_1'^2+\frac{2}{r}\delta \g_3' + \delta \g_3'',\\
0 &=& \frac{2}{r}\left(\frac{r_0}{r}-1\right) \delta \g_3'  -\frac{2 }{r^2}\delta \g_3 +\frac{2}{r^2}-\frac{3}{2} \delta v'^2\nonumber\\
&&-\frac{2}{r^2}\delta \g_1 + \delta \g_1''.
\ea

In the deterministic-view initial-point presentation, described by means of the binomial profile in the $r$ component \Eq{eq:binmeas}, one can easily find the non-trivial analytic solution
\ba
\delta \g_1 &=& \frac{\ell_*}{2}\left( \frac{r^2}{8r_0^3}-\frac{3}{4r_0} +\frac{1}{r} \right)\ln\left(1-\frac{r_0}{r}\right)\nonumber\\
 & &-\frac{\ell_*}{8r}\ln\left(\frac{r_0}{r}\right) +\frac{\ell_*}{16}\left( \frac{1}{2r_0} +\frac{r}{r_0^2}-\frac{1}{r}\right) \nonumber\\
 & &-\frac{3}{16}\frac{r_0 \ell_*}{r^2}\ln\left(\frac{r}{r_0}-1\right) \nonumber\\
\delta \g_3 &=& 1+ \frac{3\ell_*}{8}\left[\frac{1}{r_0} \ln \left(1-\frac{r_0}{r}\right) -\frac{1}{r} \ln \left(1-\frac{r}{r_0}\right) \right],  \nonumber\\
\ea
wherein we have imposed the standard solution (classical black hole in the presence of a cosmological constant) in the limit $\ell_*\to 0$. The potential $U$ is obtained from the equations of motion and is non-zero for consistency:
\ba
U & \simeq & \chi\left(1 +\sqrt{\frac{\ell_*}{r}} \right) + \mathcal{O}(\chi^2)\,.
\ea
Moreover, we can compute the new horizon radius $r_{\rm h}$. Restricting ourselves to a small deformation,
\be 
r_{\rm h} = \hat{r}_{\rm h} + \delta r,\qquad \hat{r}_{\rm h} = r_{\rm h}^{(3)},
\ee
we have
\ba
0 &=& \g_1(r_{\rm h})\simeq \g_1(\hat{r}_{\rm h}) + \delta r  \g_1'(r)\vert_{r=\hat{r}_{\rm h}}\nonumber \\
& \simeq& \tilde{\g_1}(\hat{r}_{\rm h}) + \delta \g_1(\hat{r}_{\rm h}) + \delta r \tilde{\g_1}'(r)\vert_{r=\hat{r}_{\rm h}}\nonumber  \\
& =& \delta \g_1(\hat{r}_{\rm h}) + \delta r \tilde{\g_1}'(r)\vert_{r=\hat{r}_{\rm h}},
\ea 
so that
\ba
\delta r&=& -\frac{\ell_*}{32}\left( 1+ r_0^2 \chi\right).
\ea

Once the horizon position is known, computing the Hawking temperature is straightforward:
\ba
T_{\rm H}{}_{(\Omega\neq 0)} &=& \frac{1}{4\pi}\lim_{r\to r_{\rm h}} \Bigg\vert \frac{\g_1'(r)}{v(r)}\Bigg\vert \nonumber\\
&\simeq& \frac{1}{12\pi r_{\rm h}^2}\left(1-\sqrt{\frac{\ell_*}{r_{\rm h}}}\right) \left(3 r_0 - r_{\rm h}^3 \chi \right),
\ea 
from which it is immediate to note that, when $\ell_*\to 0$, $r_{\rm h}\simeq r_{\rm h}^{(3)}$ and $T_{\rm H}{}_{(\Omega\neq 0)}\simeq T_{\rm H}^{(0)}$.\\

\subsubsection{Consequences on the evaporation time of black holes} \label{IIID}

We can repeat the same procedure to derive the evaporation time of black holes for this specific theory. Starting from \Eq{eq:evap1}, 
\be 
\frac{1}{4\pi}\int^{M_0}_0 \frac{\rmd M}{r_{\rm h}^2 T_{\rm H}^4} = \sigma \varDelta t,
\ee 
where
\ba
\frac{1}{4\pi} \int^{M_0}_0 \frac{\rmd M}{r_{\rm h}^2 T_{\rm H}^4} & \simeq & \frac{512}{5} \pi ^3 M_0 r_0 \sqrt{r_0 \ell_*} \left(\frac{5 r_0^2\chi }{4}+1\right) \nonumber\\
&& + \frac{448}{15} \pi ^3 M_0 r_0^4 \chi +\frac{64}{3} \pi ^3 M_0 r_0^2,
\ea
we can immediately derive the evaporation time and compare it with the one from the standard framework. For a test black hole with $M_0 \approx M_\odot$, one obtains again Eq.\ \Eq{eq:timepre}, the only difference being in decimals. Thus, although black-hole solutions and predictions for the Hawking temperature are inequivalent for the two values of $\Omega$ considered here, deviations with respect to standard GR are found to be of the same order.


\section{Comparison with other exotic black holes}\label{IV}

The literature on black-hole solutions in quantum-gravity approaches and alternative or modified theories of gravity is rapidly increasing \cite{DvGo,BMM,plst1,plst,gampull,casadio,garay,DGS,DLP,salvatore,olmedo,CGG,suddho}. The majority of these studies involve some departure from standard Lorentz symmetries, which are the local symmetries of GR. As discussed above, Lorentz transformations are not symmetries of spacetime when a short-distance multi-scale (in some cases, multi-fractal) behavior is considered. Despite deep conceptual and formal differences between multi-fractional gravity and other scenarios, it may be interesting to discuss, at least qualitatively, differences and similarities of black-hole solutions in some of the presently available Lorentz-violating models of gravity.

A common feature in the wide landscape of quantum gravities is a deformation of event horizons. Departures from GR at very high energy (generally at the Planck scale) affect the position of the event horizon of GR solutions. These modifications of the causal structure of black-hole spacetimes are usually tiny, being them suppressed by an ultraviolet energy scale, and become relevant only in the short-distance limit. As we shown and explained in this work, in multi-fractional gravity such effects are governed by $\ell_*$ (or, more generally, by a series of scales $\ell_n$, in the most general multi-fractional geometry \cite{rev}). For instance, a modification of the causal structure of black-hole solutions appears also in the Ho\v{r}ava--Lifshitz scenario \cite{Hor3}. In that framework, Lorentz symmetries are broken by the choice of a preferred foliation of the manifold. As a direct consequence, the theory allows for signals of arbitrarily large speed. Nonetheless, in order to preserve a notion of causality, a Killing horizon is invoked \cite{hl1,hl2,hl3}, called universal horizon. The universal horizon is defined by the condition $n^a \xi_a = 0$, where $n^a$ is the unit time-like normal vector to the space-like hypersurfaces and $\xi_a$ is a Killing vector field. Constant-time hypersurfaces can never cross it. In Ho\v{r}ava--Lifshitz gravity, black holes can have either the universal horizon alone or both a universal horizon and the standard Killing horizons of GR. In all the studied solutions, the main characteristic is that such a universal horizon lies always inside the Killing horizon. We have shown here that the horizon shrinks also in multi-fractional theories with $q$-derivatives in the initial-point presentation, while the opposite effect takes place in the final-point presentation. Moreover, regardless of the presentation choice, we have shown that multiple singularities (and horizons) come up if we take into account logarithmic oscillations of the measure. On the contrary, there are no effect on the singularity in Ho\v{r}ava--Lifshitz gravity. Such differences do not come as a surprise, since the two theories have a very similar scaling of the spectral dimension \cite{fra7} but they do not have the same symmetries.

One of the main motivations for studying alternatives to (or modifications of) GR is the desire to solve the spacetime singularities of GR. We have shown that, in general, singularity avoidance cannot be realized in the black holes of the multi-fractional theories with $q$- and weighted derivatives. This disappointing result should be compared with the achievements of both discrete-geometry theories such as loop quantum gravity and recently proposed non-local extensions of GR, where singularity avoidance is a real possibility. A class of self-consistent theories of non-local gravity, considered as a direct extension of higher-derivative models, has been recently developed in \cite{MR,CM}. In these theories, black-hole solutions are generally free of singularities \cite{BMM} and also cosmological solutions seem to enjoy finiteness at early times, in the form of a bounce \cite{CMN}. The same outcome has also been related to the possibility of having conformal invariance in the ultraviolet \cite{conform,conf1}. 

Among our findings there is the modification of the thermodynamical properties of black holes. Departures from the Hawking temperature can be encountered in the most diverse scenarios, for instance those with a generalized uncertainty principle (GUP). The GUP paradigm is an effective model that can be obtained by naively combining the quantum Heisenberg uncertainty principle with the limit on localization represented by the Schwarzschild radius. This combination gives rise to a generalized uncertainty principle. Black holes in the GUP framework have been studied mainly in \cite{Pnic,mdrBH,rainbh1,rainbh}, where it was found that sub-Planckian black holes are non-singular and obey a modified thermodynamics. In particular, the Hawking temperature scales with $M$ instead of $M^{-1}$ for such microscopic black holes. Also in multi-fractional gravity does standard thermodynamics change, but the evaporation process can either get faster or slower, depending on the presentation. Even if the behavior of $T_{\rm H}$ improves for microscopic black holes, evaporation can never be avoided in the GUP scenario. On the other hand, here we found that $T_{\rm H}$ vanishes for special values of the black-hole mass, in the case of the multi-fractional theory with $q$-derivatives in the final-point presentation.

Finally, it is interesting to compare our findings in the theory with weighted derivatives with the results in non-commutative models of gravity. Extending the idea of non-commutativity to a general covariant theory is still an open challenge, mainly due to the clash between non-commutativity (which makes explicit reference to spacetime coordinates) and invariance under diffeomorphisms. Viable ways of implementing a theory of gravitational interactions on non-commutative spaces have been proposed in the last ten years and, using different assumptions, they lead to different results. A common feature is that singularities cannot be avoided, as we found here for multi-fractional black holes. However, from our perspective, the most remarkable thing is that, in all non-commutative black-hole solutions, it is possible to relate non-commutative correction terms to the charges of classical black holes \cite{RNnonc}. In particular, the cosmological constant is generated by non-commutative effects. This is intriguingly similar to our result for the multi-fractional theory with weighted derivatives \cite{chaic,calmet}. Both a collection of small no-go theorems \cite{CR} and the fact that non-commutative gravity is less developed prevent us from establishing any rigorous connection between non-commutativity and multi-fractional theories. However, this analogy gives further support to the possibility of interpreting charges as a result of non-trivial geometries \cite{geon,cor}. Similar results are also available in modified $f(R)$ theories of gravity involving non-Riemannian spacetime properties as non-metricity \cite{olmo,bambi}. In that case, the electric charges of black holes are given a geometric interpretation \cite{olmo, bambi, olmo1}. 


\section{Conclusions}\label{V}

In this paper, we studied static and spherically symmetric black-hole solution in two different multi-fractional theories: with $q$-derivatives and with weighted derivatives. In both cases, we found departures from the Schwarzschild solution of GR. In multi-fractional gravity with $q$-derivatives, we considered two different views, one where the presentation of the measure must be fixed and another where it reflects a stochastic uncertainty. In general, the position of the event horizon changes and the Hawking temperature is modified as described in the text. In multi-fractional gravity with weighted derivatives, static and spherically symmetric black-hole solutions have a cosmological-constant term, i.e., they are Schwarzschild--de Sitter black holes. The cosmological constant arises from non-trivial geometry and it is not related to quantum fluctuations of the vacuum (we focused on classical spacetimes), in analogy with what found also in unimodular gravity.

The outlook for future research may span different directions. One is to explore in greater detail the differences and similarities with other quantum gravities discussed in Sect.\ \ref{IV}, in particular the behavior of microscopic black holes and the possibility to give black-hole charges a purely geometric pedigree. Although the information paradox problem is not resolved according to the models examined in the present work, possible non-trivial predictions related to anomalous effects may arise in quantum entanglement entropy calculations, since at smaller scales it is expected to get larger modifications. Another option could be to limit the attention to multi-fractional theories and study rotating (Kerr) black holes with the hope of finding novel phenomenology. A third possibility, however, is the following. As said in the previous section, multi-fractional gravities with $q$- and weighted derivatives do not link directly to any quantum-gravity model, at least not at the level of black holes. In this respect, the two models we studied in this paper are clearly deficient because singularities are not solved as in other quantum gravities. 

However, not all is in the negative. Even if the singularity is not avoided, we have found that it becomes non-local (in the sense of non-pointwise) in the multi-fractional theory with $q$-derivatives, in the final-point presentation. This is only one aspect of a stimulating picture. The appearance of log-periodic singularities when $r=0$ signals the breakdown of a purely metric description of spacetime, related to the discrete nature of fractal spaces at ultra-short distances. By construction, multi-fractional theories are not purely metric and the description of microscopic scales is entrusted to an exquisitely non-metric structure, encoded in the action measure and in non-standard derivative operators. The most involved, but also most realistic, realization of this anomalous integro-differential structure is the theory $T_\g$ with multi-fractional derivatives, which we have not considered here. However, we can guess the outcome from the theory with $q$-derivatives, which is an approximation of $T_\g$ \cite{rev}: most likely, black-hole singularities will not be resolved by multi-fractional derivatives. This is confirmed by a model of gravity with scale-invariant fractional derivatives, where black-hole solutions to fractional Einstein equations were constructed \cite{vac5}. Although we cannot relate those Einstein equations directly to the dynamics of $T_\g$ (the approach of \cite{vac5} is not based on a variational principle and it uses scale-invariant operators different from the multi-scale operators of $T_\g$ \cite{rev}), the basic differential structure of the dynamics is about the same. However, in the stochastic view distance uncertainties might eventually screen $T_\g$ from singularities. Our results can serve as a guiding line in anticipation of a thorough study of this alternative multi-fractional scenario.

\medskip

{\footnotesize\noindent {\bf Acknowledgments} G.C.\ is under a Ram\'on y Cajal contract and is supported by the I+D Grant FIS2014-54800-C2-2-P. D.R.F.\ is supported by the FC-15-GRUPIN-14-108 Research Grant from Principado de Asturias and partially supported by the Spanish Grant MINECO-16-FPA2015-63667-P. M.R.\ thanks Instituto de Estructura de la Materia (CSIC) for the hospitality during an early elaboration of this work. The contribution of G.C.\ and M.R.\ is based upon work from COST Action CA15117, supported by COST (European Cooperation in Science and Technology).

\noindent {\bf Open Access} This article is distributed under the terms of the Creative Commons Attribution 4.0 International License (\href{http://creativecommons.org/licenses/by/4.0/}{\cob http://creativecommons.org/licenses/by/4.0/}), which permits unrestricted use, distribution, and reproduction in any medium, provided you give appropriate credit to the original author(s) and the source, provide a link to the Creative Commons license, and indicate if changes were made.
Funded by SCOAP${}^3$.}


\begin{thebibliography}{99}
\bibitem{Matt} D.\ Mattingly, \tia{Modern tests of Lorentz invariance} \doinn{10.12942/lrr-2005-5}{Living \ Rev. \ Rel.}{8}{5}{2005} [\oarX{gr-qc/0502097}].
\bibitem{bojo2} M.\ Bojowald, \tia{Loop quantum cosmology} \doinn{10.12942/lrr-2005-11}{Living \ Rev. \ Rel.}{8}{11}{2006} [\oarX{gr-qc/0601085}].
\bibitem{gacLivRev} G.\ Amelino-Camelia, \tia{Quantum-spacetime phenomenology} \doinn{10.12942/lrr-2013-5}{Living \ Rev. \ Rel.}{16}{5}{2013} [\arX{0806.0339}].
\bibitem{qgAtom} F.\ Mercati, D.\ Maz\'on, G.\ Amelino-Camelia, J.M.\ Carmona, J.L.\ Cort\'es, J.\ Induráin, C.\ L\"ammerzahl, G.M.\ Tino, \tia{Probing the quantum-gravity realm with slow atoms} \doinn{10.1088/0264-9381/27/21/215003}{Class. \ Quantum Grav.}{27}{215003}{2010} [\arX{1004.0847}].
\bibitem{bojo1} M.\ Bojowald, \tia{Quantum cosmology: a fundamental description of the universe} \doinn{10.1007/978-1-4419-8276-6}{Lect. \ Notes \ Phys.}{3835}{1}{2011}.
\bibitem{book} G.\ Calcagni, \books{\href{10.1007/978-3-319-41127-9}{\cob Classical and Quantum Cosmology}}{Springer}{Switzerland}{2017}.
\bibitem{rev} G.\ Calcagni, \tia{Multifractional theories: an unconventional review} \doij{10.1007/JHEP03(2017)138}{JHEP}{1703}{138}{2017} [\arX{1612.05632}].
\bibitem{frc2}  G.\ Calcagni, \tia{Geometry and field theory in multi-fractional spacetime} \doij{10.1007/JHEP01(2012)065}{JHEP}{1201}{065}{2012} [\arX{1107.5041}].
\bibitem{frc11} G.\ Calcagni, \tia{Multi-scale gravity and cosmology} \doij{10.1088/1475-7516/2013/12/041}{JCAP}{1312}{041}{2013} [\arX{1307.6382}].
\bibitem{trtls} G.\ Calcagni, \tia{ABC of multi-fractal spacetimes and fractional sea turtles} \doin{10.1140/epjc/s10052-016-4021-0}{Eur.\ Phys.\ J.}{C}{76}{181}{2016} [\arX{1602.01470}].
\bibitem{first} G.\ Calcagni, \tia{Multiscale spacetimes from first principles} \doin{10.1103/PhysRevD.95.064057}{Phys.\ Rev.}{D}{95}{064057}{2017} [\arX{1609.02776}].
\bibitem{SKM}   S.\ Samko, A.\ Kilbas, O.\ Marichev, \book{Fractional Integrals and Derivatives}{Gordon and Breach}{New York}{U.S.A.}{1993}.
\bibitem{GoMa}  R.\  Gorenflo, F.\ Mainardi, \tia{Fractional calculus} \proc{Fractals and Fractional Calculus in Continuum Mechanics}{A.\ Carpinteri, F.\ Mainardi}{Springer}{Heidelberg}{Germany}{1997}. 
\bibitem{KST}   A.A.\ Kilbas, H.M.\ Srivastava, J.J.\ Trujillo, \book{Theory and Applications of Fractional Differential Equations}{Elsevier}{Amsterdam}{The Netherlands}{2006}.
\bibitem{Die10} K.\ Diethelm, \book{\href{10.1007/978-3-642-14574-2}{\cob The Analysis of Fractional Differential Equations}}{Springer}{Heidelberg}{Germany}{2010}.
\bibitem{frc1}  G\ Calcagni, \tia{Geometry of fractional spaces} \doinn{10.4310/ATMP.2012.v16.n2.a5}{Adv.\ Theor.\ Math.\ Phys.}{16}{549}{2012} [\arX{1106.5787}].
\bibitem{ElT2}  R.A.\ El-Nabulsi, D.F.M.\ Torres, \tia{Fractional actionlike variational problems} \doinn{10.1063/1.2929662}{J.\ Math. Phys.}{49}{053521}{2008} [\arX{0804.4500}]. 
\bibitem{Jac09} F.H.\ Jackson, \tia{On $q$-functions and a certain difference operator} \doinn{10.1017/S0080456800002751}{Trans.\ R.\ Soc.\ Edin.}{46}{253}{1909}.
\bibitem{Tsa09} C.\ Tsallis, \book{\href{10.1007/978-0-387-85359-8}{\cob Introduction to Nonextensive Statistical Mechanics}}{Springer}{New York}{U.S.A.}{2009}.
\bibitem{minlen} G.\ Amelino-Camelia, G.\ Calcagni, M.\ Ronco, \tia{Imprint of quantum gravity in the dimension and fabric of spacetime} \arX{1705.04876}.
\bibitem{CR}  G.\ Calcagni, M.\ Ronco, \tia{Deformed symmetries in noncommutative and multifractional spacetimes} \doin{10.1103/PhysRevD.95.045001 }{Phys.\ Rev.}{D}{95}{045001}{2017} [\arX{1608.01667}].
\bibitem{Calcagni:2016ofu} G.\ Calcagni, S.\ Kuroyanagi, S.\ Tsujikawa, \tia{Cosmic microwave background and inflation in multi-fractional spacetimes} \doij{10.1088/1475-7516/2016/08/039}{JCAP}{1608}{039}{2016} [\arX{1606.08449}].
\bibitem{infopar1} J.D.\ Bekenstein, \tia{Black holes and entropy} \doin{10.1103/PhysRevD.7.2333}{Phys.\ Rev.}{D}{7}{2333}{1973}.
\bibitem{infopar2} S.W.\ Hawking, \tia{Particle creation by black holes} \doinn{10.1007/BF02345020}{Commun.\ Math.\ Phys.}{46}{206}{1975}.
\bibitem{infopar3} S.B.\ Giddings, \tia{Black hole information, unitarity, and nonlocality} \doin{10.1103/PhysRevD.74.106005 }{Phys.\ Rev.}{D}{74}{106005}{2006} [\arX{ hep-th/0605196}].
\bibitem{infopar4} W.G.\ Unruh, \tia{Notes on black hole evaporation} \doin{10.1103/PhysRevD.14.870}{Phys.\ Rev.}{D}{14}{870}{1976}.
\bibitem{infopar5} A.\ Strominger, C.\ Vafa, \tia{Microscopic origin of the Bekenstein--Hawking entropy} \doin{10.1016/0370-2693(96)00345-0 }{Phys.\ Lett.}{B}{379}{99}{1996} [\oarX{hep-th/9601029}].
\bibitem{infopar6} M.\ Bojowald, \tiaq{Information loss, made worse by quantum gravity?} \doinn{10.3389/fphy.2015.00033}{Front.\ Phys.}{3}{33}{2015} [\arX{1409.3157}].	
\bibitem{CalcagniArzano} M.\ Arzano, G.\ Calcagni, \tia{What gravity waves are telling about quantum spacetime} \doin{10.1103/PhysRevD.93.124065}{Phys.\ Rev.}{D}{93}{124065 }{2016} [\arX{1604.00541}].
\bibitem{Arzano}  M.\  Arzano, \tia{Black hole entropy, log corrections and quantum ergosphere} \doin{10.1016/j.physletb.2006.02.020}{Phys.\ Lett.}{B}{634}{536}{2006} [\oarX{gr-qc/0512071}]. 
\bibitem{chi1}  J.M.\ Bardeen, \tia{Black holes do evaporate thermally} \doinn{10.1103/PhysRevLett.46.382}{Phys.\ Rev.\ Lett}{46}{382}{1981}.
\bibitem{Kagramanova:2006ax}  V.\ Kagramanova, J.\ Kunz, C.\ L\"ammerzahl, \tia{Solar system effects in Schwarzschild--de Sitter spacetime} \doin{10.1016/j.physletb.2006.01.069}{Phys.\ Lett.}{B}{634}{251}{2011} [\oarX{gr-qc/0602002}].
\bibitem{Adams}  F.C.\ Adams, M.\ Mbonye, G.\ Laughlin, \tia{Possible effects of a cosmological constant on black hole evolution} \doin{10.1016/S0370-2693(99)00174-4}{Phys.\ Lett.}{B}{450}{339}{1999} [\oarX{astro-ph/9902118}].
\bibitem{cosmocon1} T.\ Padmanabhan, \tia{Cosmological constant: the weight of the vacuum} \doinn{10.1016/S0370-1573(03)00120-0}{Phys.\ Rep.}{380}{235}{2003} [\oarX{hep-th/0212290}].
\bibitem{geon} J.A.\ Wheeler, \tia{Geons} \doinn{10.1103/PhysRev.97.511}{Phys.\ Rev.}{97}{511}{1955}.
\bibitem{Alv1}  E.\ Álvarez, S.\ González-Martín, M.\ Herrero-Valea, \tia{Quantum corrections to unimodular gravity} \doij{10.1007/JHEP08(2015)078}{JHEP}{1508}{078}{2015} [\arX{1505.01995}].
\bibitem{Smolin}  L.\ Smolin, \tia{Quantization of unimodular gravity and the cosmological constant problems} \doin{10.1103/PhysRevD.80.084003}{Phys.\ Rev.}{D}{80}{084003 }{2009} [\arX{0904.4841}].
\bibitem{Alv2}  E.\ Álvarez, M.\ Herrero-Valea, \tia{Unimodular gravity with external sources} \doij{10.1088/1475-7516/2013/01/014}{JCAP}{1301}{014}{2013} [\arX{1209.6223}].
\bibitem{DamNg}  J.J.\ van der Bij, H.\ van Dam, Y.J.\ Ng, \tia{The exchange of massless spin-two particles} \doin{10.1016/0378-4371(82)90247-3}{Physica}{A}{116}{307}{1982}.
\bibitem{nat1}  X.-B.\ Wu {et al.}, \tia{An ultraluminous quasar with a twelve-billion-solar-mass black hole at redshift 6.30} \doinn{10.1038/nature14241}{Nature}{518}{515}{2015} [\arX{1502.07418}].
\bibitem{frc13} G.\ Calcagni, G.\ Nardelli, D.\ Rodr\'iguez-Fern\'andez, \tia{Standard Model in multiscale theories and observational constraints} \doin{10.1103/PhysRevD.94.045018}{Phys.\ Rev.}{D}{94}{045018}{2016} [\arX{1512.06858}].
\bibitem{frc3}  G.\ Calcagni, G.\ Nardelli, \tia{Momentum transforms and Laplacians in fractional spaces} \doinn{10.4310/ATMP.2012.v16.n4.a5}{Adv.\ Theor.\ Math.\ Phys.}{16}{1315}{2012} [\arX{1202.5383}].
\bibitem{DvGo}  G.\ Dvali, C.\ Gomez, \tia{Black hole's quantum N-portrait} \doinn{10.1002/prop.201300001}{Fortschr.\ Phys.}{61}{742}{2013} [\arX{1112.3359}].
\bibitem{BMM}   C.\ Bambi, D.\ Malafarina, L.\ Modesto, \tia{Terminating black holes in quantum gravity} \doin{10.1140/epjc/s10052-014-2767-9}{Eur.\ Phys.\ J.}{C}{74}{2767}{2014} [\arX{1306.1668}]. 
\bibitem{plst1} C.\ Rovelli, F.\ Vidotto, \tia{Planck stars} \doin{10.1142/S0218271814420267}{Int.\ J.\ Mod.\ Phys.}{D}{23}{1442026}{2014} [\arX{1401.6562}].
\bibitem{plst} T.\ De Lorenzo, C.\ Pacilio, C.\ Rovelli, S.\ Speziale, \tia{On the effective metric of a Planck star} \doinn{10.1007/s10714-015-1882-8}{Gen.\ Relat.\ Gravit.}{47}{41}{2014} [\arX{1412.6015]}.
\bibitem{gampull} R.\ Gambini, J.\ Pullin, \tia{Loop quantization of the Schwarzschild black hole} \doinn{10.1103/PhysRevLett.110.211301}{Phys.\ Rev.\ Lett.}{110}{211301}{2013} [\arX{1302.5265}].
\bibitem{casadio} R.\ Casadio, A.\ Giugno, O.\ Micu, A.\ Orlandi, \tia{Black holes as self-sustained quantum states, and Hawking radiation} \doin{10.1103/PhysRevD.90.084040}{Phys.\ Rev.}{D}{90}{084040}{2014} [\arX{1405.4192}].
\bibitem{garay} C.\ Barceló, R.\ Carballo-Rubio, L.J.\ Garay, \tiaq{Where does the physics of extreme gravitational collapse reside?} \doinn{10.3390/universe2020007}{Universe}{2}{7}{2016} [\arX{1510.04957}].
\bibitem{DGS} T.\ De Lorenzo, A.\ Giusti, S.\ Speziale, \tia{Non-singular rotating black hole with a time delay in the center} \doinn{10.1007/s10714-016-2105-7}{Gen.\ Relat.\ Gravit.}{48}{1}{2016} [\arX{1510.08828}].
\bibitem{DLP} T.\ De Lorenzo, A.\ Perez, \tia{Improved black hole fireworks: Asymmetric black-hole-to-white-hole tunneling scenario} \doin{10.1103/PhysRevD.93.124018}{Phys.\ Rev.}{D}{93}{124018}{2016} [\arX{1512.04566}].
\bibitem{salvatore} A.\ Addazi, S.\ Capozziello,  \tia{The fate of Schwarzschild--de Sitter black holes in $f(R)$ gravity} \doin{10.1142/S0217732316500541}{Mod.\ Phys.\ Lett.}{A}{31}{1650054}{2016} [\arX{1602.00485}].
\bibitem{olmedo} J.\ Olmedo, \tia{Brief review on black hole loop quantization} \doinn{10.3390/universe2020012}{Universe}{2}{12}{2016} [\arX{1606.01429}].
\bibitem{CGG} R.\ Casadio, A.\ Giugno, A.\ Giusti, \tia{Matter and gravitons in the gravitational collapse} \doin{10.1016/j.physletb.2016.10.058}{Phys.\ Lett.}{B}{763}{337}{2016} [\arX{1606.04744}]. 
\bibitem{suddho} M.\ Bojowald, S.\ Brahma, \tia{Signature change in 2-dimensional black-hole models of loop quantum gravity} \arX{1610.08850}.
\bibitem{Hor3}  P.\ Ho\v{r}ava, \tia{Spectral dimension of the universe in quantum gravity at a Lifshitz point} \doinn{10.1103/PhysRevLett.102.161301}{Phys.\ Rev.\ Lett.}{102}{161301}{2009} [\arX{0902.3657}]. 
\bibitem{hl1} T.P.\ Sotiriou, I.\ Vega, D.\ Vernieri, \tia{Rotating black holes in three-dimensional Ho\v{r}ava gravity} \doin{10.1103/PhysRevD.90.044046}{Phys.\ Rev.}{D}{90}{044046}{2014} [\arX{1405.3715}].
\bibitem{hl2}     M.\ Saravani, N.\ Afshordi, R.B.\ Mann,  \tia{Dynamical emergence of universal horizons during the formation of black holes} \doin{10.1103/PhysRevD.89.084029}{Phys.\ Rev.}{D}{89}{084029}{2014} [\arX{1310.4143}].
\bibitem{hl3} K.\ Lin, E.\ Abdalla, R.G.\ Cai, A.\ Wang, \tia{Universal horizons and black holes in gravitational theories with broken Lorentz symmetry} \doin{10.1142/S0218271814430044}{Int.\ J.\ Mod.\ Phys.}{D}{23}{1443004}{2014} [\arX{1408.5976}].
\bibitem{fra7}  G.\ Calcagni, \tia{Multifractional spacetimes, asymptotic safety and Ho\v{r}ava--Lifshitz gravity} \doin{10.1142/S0217751X13500929}{Int.\ J.\ Mod.\ Phys.}{A}{28}{1350092}{2013} [\arX{1209.4376}].
\bibitem{MR} L.\ Modesto, L.\ Rachwał, \tia{Universally finite gravitational and gauge theories} \doin{10.1016/j.nuclphysb.2015.09.006}{Nucl.\ Phys.}{B}{900}{147}{2015} [\arX{0905.4949}].
\bibitem{CM}  G.\ Calcagni, L.\ Modesto, \tia{Nonlocal quantum gravity and M-theory} \doin{10.1103/PhysRevD.91.124059}{Phys.\ Rev.}{D}{91}{124059}{2015} [\arX{1404.2137}].
\bibitem{CMN}   G.\ Calcagni, L.\ Modesto, P.\ Nicolini, \tia{Super-accelerating bouncing cosmology in asymptotically-free non-local gravity} \doin{10.1140/epjc/s10052-014-2999-8}{Eur.\ Phys.\ J.}{C}{74}{2999}{2014} [\arX{1306.5332}].
\bibitem{conform} C.\ Bambi, L.\ Modesto, S.\ Porey, L.\ Rachwał, \tia{Black hole evaporation in conformal gravity} \arX{1611.05582}.
\bibitem{conf1} B.\ Boisseau, H.\ Giacomini, D.\ Polarski, \tia{Bouncing universes in scalar-tensor gravity around conformal invariance} \doij{10.1088/1475-7516/2016/05/048}{JCAP}{1605}{048}{2016} [\arX{1603.06648}].
\bibitem{Pnic} P.\ Nicolini, A.\ Smailagic, E.\ Spallucci, \tia{Noncommutative geometry inspired Schwarzschild black hole} \doin{10.1016/j.physletb.2005.11.004}{Phys.\ Lett.}{B}{632}{547}{2006} [\oarX{gr-qc/0510112}].
\bibitem{mdrBH}  A.D.\ Kamali, P.\ Aspoukeh, \tia{Corrections to the Hawking tunneling radiation from MDR} \doinn{10.1007/s10773-016-3072-1}{Int. \ J. \ Theor.\ Phys.}{55}{4492}{2016}.
\bibitem{rainbh1}  A.\ Farag Ali, M.\  Faizal, M.M.\ Khalil, \tia{Absence of black holes at LHC due to gravity's rainbow} \doin{10.1016/j.physletb.2015.02.065}{Phys.\ Lett.}{B}{743}{295}{2015} [\arX{1410.4765}].
\bibitem{rainbh} J.\ Tao, P.\ Wang, H.\ Yang, \tia{Free-fall frame black hole in gravity's rainbow} \doin{10.1103/PhysRevD.94.064068}{Phys.\ Rev.}{D}{94}{064068}{2016} [\arX{1602.08686}].
\bibitem{RNnonc} R.\ Bufalo, A.\ Tureanu, \tia{Analogy between the Schwarzschild solution in a noncommutative gauge theory and the Reissner--Nordström metric} \doin{10.1103/PhysRevD.92.065017}{Phys.\ Rev.}{D}{92}{065017}{2015} [\arX{1410.8661}].
\bibitem{chaic} M.\ Chaichian, A.\ Tureanu, G.\ Zet, \tia{Corrections to Schwarzschild solution in noncommutative gauge theory of gravity} \doin{10.1016/j.physletb.2008.01.029}{Phys.\ Lett.}{B}{660}{573}{2015} [\arX{0710.2075}].
\bibitem{calmet} X.\ Calmet, \tia{Cosmological constant and noncommutative spacetime} \doinn{10.1209/0295-5075/77/19002}{Europhys.\ Lett.}{77}{19902}{2007} [\arX{hep-th/0510165}].
\bibitem{cor} A.\ Corichi, P.\ Singh, \tia{Geometric perspective on singularity resolution and uniqueness in loop quantum cosmology} \doin{10.1103/PhysRevD.80.044024}{Phys.\ Rev.}{D}{80}{044024}{2009} [\arX{0905.4949}].
\bibitem{olmo}  G.J.\ Olmo, D.\ Rubiera-García, A.\ Sánchez-Puente, \tia{Geodesic completeness in a wormhole spacetime with horizons} \doin{10.1103/PhysRevD.92.044047}{Phys.\ Rev.}{D}{92}{044047}{2015} [\arX{1508.03272}].
\bibitem{bambi}  C.\ Bambi, A.\ Cardenas-Avendano, G.J.\ Olmo, D.\ Rubiera-García, \tia{Wormholes and nonsingular spacetimes in Palatini $f(R)$ gravity} \doin{10.1103/PhysRevD.93.064016}{Phys.\ Rev.}{D}{93}{064016}{2016} [\arX{1511.03755}].
\bibitem{olmo1}  C.\ Bambi, A.\ Cardenas-Avendano, G.J.\ Olmo, D.\ Rubiera-García, \tia{The quantum, the geon, and the crystal} \doin{10.1142/S0218271815420134}{Int.\ J.\ Mod.\ Phys.}{D}{24}{1542013}{2015} [\arX{1507.07777}].
\bibitem{vac5}  S.I.\ Vacaru, \tia{Fractional dynamics from Einstein gravity, general solutions, and black holes} \doinn{10.1007/s10773-011-1010-9}{Int.\ J.\ Theor.\ Phys.}{51}{1338}{2012} [\arX{1004.0628}].
\end{thebibliography}
\end{document}